\documentclass[]{aa}
\usepackage[varg]{txfonts}
\usepackage{pdflscape}
\usepackage{adjustbox}
\usepackage{xcolor}
\usepackage{placeins}
\usepackage{lipsum}

\begin{document}

\title{Dynamical masses for two M1 + mid-M dwarf binaries monitored during the SPHERE-SHINE survey}


\author{Beth A. Biller\inst{\ref{ROE}, \ref{CES}, \ref{MPIA}} \and Antoine Grandjean \inst{\ref{Grenoble}} \and Sergio Messina\inst{\ref{Catania}} \and Silvano Desidera\inst{\ref{INAF}} \and Philippe Delorme\inst{\ref{Grenoble}} \and Anne-Marie Lagrange\inst{\ref{Grenoble}, \ref{LESIA}, \ref{IMCCE}} \and Franz-Josef Hambsch\inst{\ref{AAVSO}, \ref{VVS}} \and Dino Mesa\inst{\ref{INAF}} \and Markus Janson\inst{\ref{Stockholm}, \ref{MPIA}} \and Raffaele Gratton\inst{\ref{INAF}} \and Valentina D'Orazi\inst{\ref{INAF}} \and Maud Langlois\inst{\ref{Lyon}, \ref{LAM}} \and Anne-Lise Maire\inst{\ref{Liege}, \ref{MPIA}} \and Joshua Schlieder\inst{\ref{Goddard}, \ref{MPIA}} \and Thomas Henning\inst{\ref{MPIA}} \and Alice Zurlo\inst{\ref{UDP1}, \ref{UDP2}, \ref{LAM}} \and Janis Hagelberg\inst{\ref{Geneva}} \and S. Brown\inst{\ref{MPIA}} \and C. Romero\inst{\ref{Grenoble}, \ref{ESOChile}} \and Micka\"el Bonnefoy\inst{\ref{Grenoble}} \and Gael Chauvin\inst{\ref{Grenoble}, \ref{FCLA}} \and Markus Feldt\inst{\ref{MPIA}} \and Michael Meyer\inst{\ref{Michigan}, \ref{ETH}} \and Arthur Vigan\inst{\ref{LAM}} \and A. Pavlov\inst{\ref{MPIA}} \and C. Soenke\inst{\ref{ESO}} \and D. LeMignant\inst{\ref{LAM}} \and A. Roux\inst{\ref{Grenoble}}}

\offprints{B.Biller, \email{bb@roe.ac.uk}}
\institute{SUPA, Institute for Astronomy, University of Edinburgh, Blackford Hill, Edinburgh EH9 3HJ, UK \label{ROE} \and
Centre for Exoplanet Science, University of Edinburgh, Edinburgh, UK \label{CES} \and
Max Planck Institute for Astronomy, K\"onigstuhl 17, D-69117 Heidelberg, Germany \label{MPIA} \and
Univ. Grenoble Alpes, CNRS, IPAG, 38000 Grenoble, France \label{Grenoble} \and
INAF-Catania Astrophysical Observatory, via S.Sofia, 78 I-95123 Catania, Italy \label{Catania} \and
INAF - Osservatorio Astronomico di Padova, Vicolo della Osservatorio 5, 35122, Padova, Italy \label{INAF} \and
LESIA, Observatoire de Paris, Universit\'e PSL, CNRS, Sorbonne
Universit\'e, Universit\'e de Paris, 5 place Jules Janssen, 92195
Meudon, France \label{LESIA} \and
IMCCE – Observatoire de Paris, 77 Avenue Denfert-Rochereau,
75014 Paris, France \label{IMCCE} \and
American Association of Variable Star Observers (AAVSO), 49 Bay State
Rd., Cambridge, MA 02138, USA \label{AAVSO} \and
Vereniging Voor Sterrenkunde (VVS), Oostmeers 122 C, 8000 Brugge, Belgium \label{VVS} \and
Institutionen f\"or astronomi, Stockholms Universitet, Stockholm, Sweden \label{Stockholm} \and
CRAL, UMR 5574, CNRS, Universit\'e de Lyon, ENS, 9 avenue
Charles Andr\'e, 69561 Saint Genis Laval Cedex, France \label{Lyon} \and
Aix Marseille Univ, CNRS, CNES, LAM, Marseille, France \label{LAM} \and
STAR Institute, Universit\'e de Li\`ege, All\'ee du Six Ao\^ut 19c, B-4000
Li\`ege, Belgium \label{Liege} \and
Exoplanets and Stellar Astrophysics Laboratory, NASA Goddard Space Flight Center, Greenbelt, MD, United States \label{Goddard} \and
N\'ucleo de Astronom\'ia, Facultad de Ingenier\'ia y Ciencias, Universidad Diego Portales, Av. Ejercito 441, Santiago, Chile \label{UDP1} \and
Escuela de Ingenier\'ia Industrial, Facultad de Ingenier\'ia y Ciencias, Universidad Diego Portales, Av. Ejercito 441, Santiago, Chile \label{UDP2} \and
Geneva Observatory, University of Geneva, 51 ch. Pegasi, CH-1290 Versoix, Switzerland \label{Geneva} \and
ESO Vitacura, Alonso de C\'ordova 3107, Vitacura, Casilla 19001, Santiago de Chile, Chile \label{ESOChile} \and
Unidad Mixta Internacional Franco-Chilena de Astronom\'ia, CNRS/INSU UMI 3386 and Departamento de Astronomía, Universidad de Chile, Casilla 36-D, Santiago, Chile\label{FCLA} \and
Department of Astronomy, University of Michigan, Ann Arbor, Ann Arbor, MI \label{Michigan} \and
Institute for Particle Physics and Astrophysics, ETH Zurich, Wolfgang-Pauli-Strasse 27, 8093 Zurich, Switzerland \label{ETH} \and
European Southern Observatory (ESO), Karl-Schwarzschild-Str. 2, 85748 Garching, Germany \label{ESO}}

\date{}

\abstract{We present orbital fits and dynamical masses for HIP 113201AB and HIP 36985AB, two M1 + mid-M dwarf binary systems monitored as part of the SPHERE SHINE survey.  To robustly determine the age of both systems via gyrochronology, we undertook a photometric monitoring campaign for HIP 113201 and for GJ 282AB, the two wide K star companions to HIP 36985, using the 40 cm Remote Observatory Atacama Desert (ROAD) telescope.  Based on this monitoring and gyrochronological relationships, we adopt ages of 1.2$\pm$0.1 Gyr for HIP 113201AB and 750$\pm$100 Myr for HIP 36985AB. These systems are sufficiently old that we expect that all components of these binaries will have reached the main sequence. 
To derive dynamical masses for all components of the HIP 113201AB and HIP 36985AB systems, we used parallel-tempering Markov Chain Monte Carlo sampling  to fit a combination of radial velocity, direct imaging, and Gaia and Hipparcos astrometry.
Fitting the direct imaging and radial velocity data for HIP 113201 yields a primary mass of 0.54$\pm$0.03 M$_{\odot}$, fully consistent with its M1 spectral type, and a secondary mass of 0.145$\pm$ M$_{\odot}$.  The secondary masses derived with and without including Hipparcos/Gaia data are all considerably more massive than the 0.1 M$_{\odot}$ estimated mass from the photometry of the companion.  Thus, the dynamical impacts of this companion suggest that it is more massive than expected from its photometry. 
An undetected brown dwarf companion to HIP 113201B could be a natural explanation for this apparent discrepancy.  At an age $>$1 Gyr, a 30 M$_{Jup}$ companion to HIP 113201B would make a negligible ($<$1$\%$) contribution to the system luminosity, but could have strong dynamical impacts. Fitting the direct imaging, radial velocity, and Hipparcos/Gaia proper motion anomaly for HIP 36985AB, we find a primary mass of 0.54$\pm$0.01 M$_{\odot}$ and a secondary mass of 0.185$\pm$0.001 M$_{\odot}$ which agree well with photometric estimates of component masses, the masses estimated from $M_{K}$-- mass relationships for M dwarf stars, and previous dynamical masses in the literature.
} 


\titlerunning{Dynamical masses for two M1 + mid-M dwarf binaries}
\authorrunning{B.A. Biller et al.}
\maketitle 

\section{Introduction}

Dynamical masses for components of stellar binaries are critical to benchmark the computational models used to estimate the masses and radii of isolated objects based on their luminosities.  While abundant dynamical masses for high-mass and solar mass stellar binaries are reported in the literature, fewer dynamical masses are available for M dwarfs with masses $<$0.3 M$_{\odot}$ and ultracool dwarfs, especially at young ($<$1 Gyr) ages. 
In the last decade, empirical mass-luminosity relationships have been constructed for late K dwarf and M dwarf stars. \citet{Benedict2016} construct a mass-luminosity relationship for M dwarf stars via HST orbital monitoring of 47 stars with masses from 0.08 - 0.62 M$_{\odot}$. \citet{Mann2019} derive an empirical Mass-Luminosity-Metallicity Relation for stellar masses between 0.075 and 0.7 M$_{\odot}$ by measuring total system masses from 62 nearby M-dwarf binaries.  

Many dynamical masses in the literature via direct imaging of visual binaries are in fact total masses for the system.  
Direct imaging of the motion of a companion relative to its primary must be combined with either radial velocity (henceforth RV) or absolute astrometry of the primary in order to determine individual component masses. Combining CFHT absolute astrometry with adaptive-optics assisted resolved imaging relative to the primary, \citet{Dupuy2017} present 38 individual masses for ultracool dwarfs (spectral type $\geq$M6), ranging from 30 to 115 M$_{Jup}$, the largest such sample for substellar objects and ultracool dwarfs in the literature.
A number of low mass and substellar companions have been directly imaged after their presence was indirectly noted via a radial velocity trend \citep[e.g.~][]{Crepp2016}. For companions to relatively bright stars, the combination of Gaia DR2 or EDR3 proper motions with Hipparcos proper motions can provide additional absolute astrometry orbital constraints \citep{Kervella2019}.  Only in the last few years have these techniques been fully combined to determine dynamical masses for a range of stars, substellar objects and exoplanets \citep{Bowler2018, Calissendorff2018, Dieterich2018, Snellen2018, Grandjean2019, Dupuy2019, Brandt2019, Brandt2021b}.  Here we present dynamical mass determinations combining radial velocity, direct imaging, and Gaia /Hipparcos data for all components of the HIP 113201 and HIP 36985 binary systems.  

HIP 113201AB and HIP 36985AB are two M1 + mid-M binary systems.  The existence of these binaries were first suggested via their proper motion anomalies \citep{Kervella2019} -- slight differences between the proper motions of the primaries as measured by Hipparcos and Gaia suggested the presence of an additional component in each system.  Both binaries have since been confirmed via direct imaging and RV observations.  HIP 113201B was discovered by \citet{Bonavita21} in the course of the SPHERE/SHINE survey \citep{Chauvin2017b}, a $\sim$500 star survey targeting young stellar systems using the SPHERE planet-finding camera \citep{Beuzit2019} in order to directly image young giant exoplanets \citep{Desidera2021,Langlois2021,Vigan2020}.  The HIP 113201AB system also has extensive RV monitoring with HARPS. 
For HIP 36985B, \citet{Kammerer2019} first imaged the companion via VLT/NaCo kernel phase imaging and direct VLT/NaCo imaging. \citet{Grandjean2020} confirmed that HIP 36985B was bound and covered a significant fraction of the companion's orbit via HARPS RV observations.  \citet{Bonavita21} provided additional epochs of VLT-SPHERE imaging of this companion and \citet{Baroch2021} has recently published additional CARMENES and FEROS RV observations. With preliminary dynamical masses $<$0.2 M$_{\odot}$ \citep{Bonavita21, Baroch2021}, HIP 113201B and HIP 36985B are two of the lowest mass stellar companions imaged as part of the SPHERE/SHINE survey.

In Section~\ref{sec:properties}, we summarize the properties of both stellar systems studied here, and present updated ages for the systems.  In Section~\ref{sec:mcmc}, we describe the parallel-tempering Markov chain Monte Carlo (PT-MCMC) code used to fit orbits for both systems.   Section~\ref{sec:orbits} covers results from our PT-MCMC fits to both systems.  We compare our derived dynamical masses for all components of both systems to existing mass estimates in Section~\ref{sec:discussion} and summarize our conclusions in Section~\ref{sec:conclusions}.

\section{Properties of the Primary Stars\label{sec:properties}}

\subsection{HIP 113201 / GJ 4303 / BPM 28050}
\citet{Gaidos2014} fit the optical spectrum of HIP 113201 and find a best fit spectral type of M1, with best fit values of T$_\mathrm{eff}$ = 3693$\pm$82 K, luminosity = 0.038$\pm$0.009 L$_{\odot}$, radius = 0.49 $\pm$0.05 R$_{\odot}$, and mass = 0.52$\pm$0.07 M$_{\odot}$.  \citet{Hawley1996} find a similar but slightly earlier M0.5 spectral type from low-resolution optical spectroscopy using the CTIO 1.5 m telescope.  We used the $M_{K}$-mass relationship for stars with masses between 0.075 and 0.7 M$_{\odot}$ derived from orbital fits to 62 nearby binaries by \citet{Mann2019} to estimate a mass of 0.52$\pm$0.01 M$_{\odot}$ for HIP 113201, matching the \citet{Gaidos2014} spectroscopic mass estimate.  Stellar properties are presented in Table~\ref{tab:primaries}. 

HIP 113201 was included in the SPHERE SHINE survey sample because its kinematics are a reasonable match to both the Tucana-Horologium and $\beta$ Pic moving groups.  However, using a variety of age diagnostics, the star appears to be older than comparable spectral type members in both groups.  Its weak H-alpha emission coupled with a lack of lithium absorption suggest an age equal to or greater than the Hyades -- in other words, somewhat younger than typical field stars, but not $<$300 Myr.  

\begin{figure*}
\includegraphics[scale=0.6, angle=90]{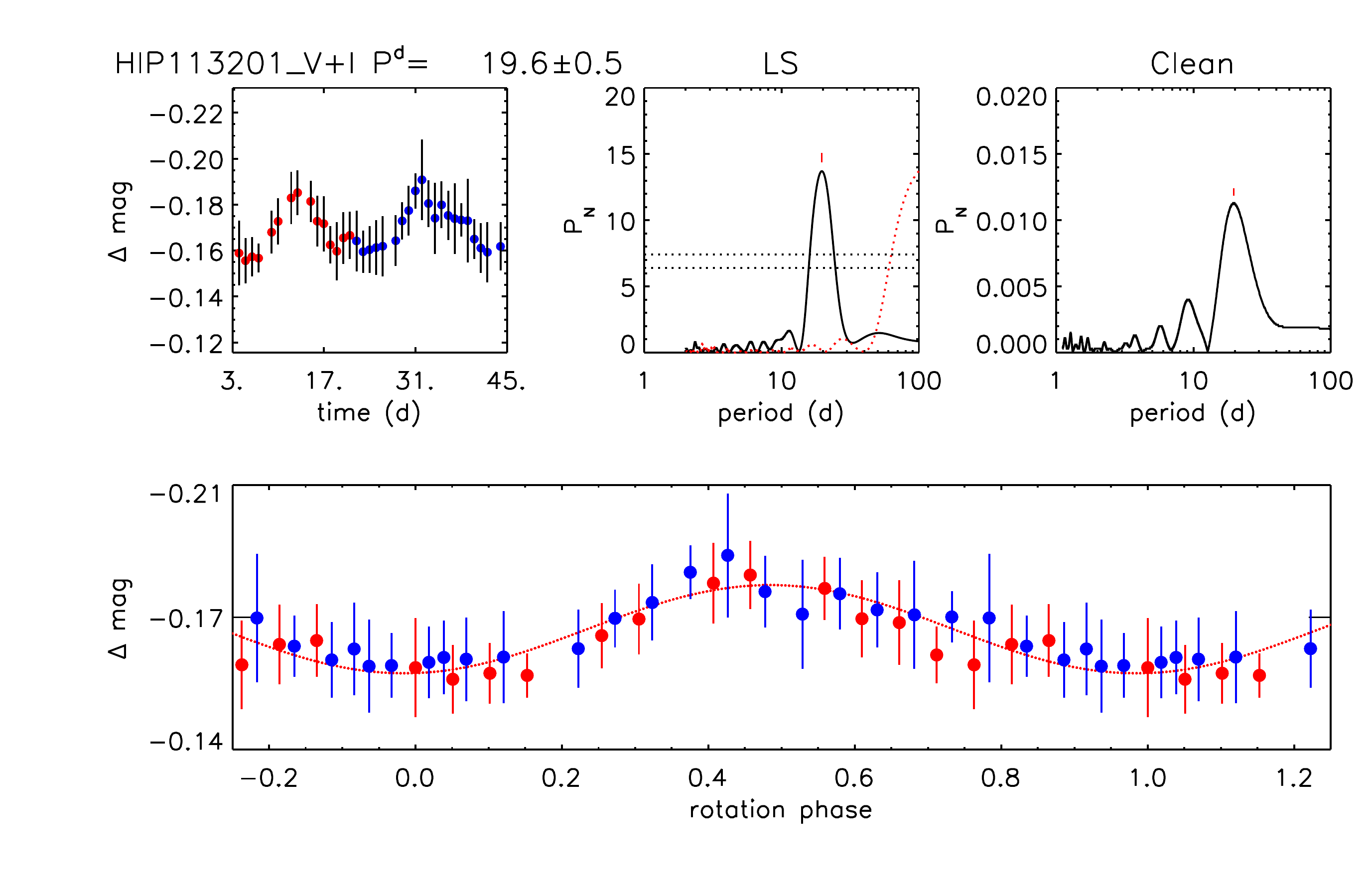}
\caption{Summary of periodogram analysis for HIP\,113201. Top-left panel: combined V+I-band magnitudes versus HJD (-2457670). Measurements on each night were taken in both V band and I band; the average of the two bands is plotted here.  Top-middle panel: Lomb-Scargle periodogram. The horizontal dotted lines show the 99.9\% and 98\% confidence levels, whereas the red dotted line shows the spectral window function. Top-right panel: the Clean periodogram. Bottom panel: the lightcurve phased with the 19.6-d rotation period.
Red circles are data points from the first half of the observations; blue circles are data points from the second half of the observations.
The red dotted line is a sinusoidal fit to the phased light curve.}
\label{fig:periodogram_HIP113201}       
\end{figure*}

To robustly determine the age via gyrochronology, we undertook a photometric monitoring campaign in October / November 2016 at the Remote Observatory Atacama Desert (ROAD) located in the
Atacama Desert close to the town of San Pedro de Atacama,
Chile. The telescope is a 40 cm
f /6.8 Optimized Dall-Kirkham (ODK),
equipped with a 4K$\times$4K pixel FLI ML16803 CCD camera (9 $\mu$m
pixel size) with a 40$^{\prime}$$\times$40$^{\prime}$ FoV and BVI filters.
We observed HIP113201 for 36 almost consecutive nights spanning a time interval of 40 days. We collected a total of 179 frames in V filter and 179 frames in I filter (generally one telescope pointing per night with five consecutive frames in each filter that were averaged to get one average magnitude per night and corresponding $\sigma$).  We used aperture photometry to extract the magnitudes and computed differential magnitudes of HIP113201 with respect to an ensemble comparison consisting of three non variable stars (TYC\,8453\,806\,1; CD$-$539254; TYC\,8453\,856\,1). The photometric precision is 0.007 mag in both I and V filters.
We performed Lomb-Scargle (LS) and CLEAN periodogram analysis on the average data for the V, the I, and the (V+I)/2 time series. The latter provides the most precise results, which are summarized in Fig.~\ref{fig:periodogram_HIP113201}.  In the periodogram of V-filter data we find a period of 19.9$\pm$0.7 days and a light curve amplitude 0.03 mag.  In the periodogram of I-filter data we find a period of 19.2$\pm$0.7 days and a light curve amplitude 0.02 mag.  In the periodogram of the  combined light curve, we find a period of 19.6$\pm$0.5 days, which we adopt as the photometric rotation period for this star.  The rotation period is highly significant (confidence level $>$ 99\%) with coverage over multiple rotation periods. The V and I magnitude variations are correlated with linear Pearson correlation coefficient 0.6 and significance 99.9\%, suggesting that the photometric variability is dominated by either cool spots or hot spots (faculae).

HIP 113201 was also monitored by the Transiting Exoplanet Survey Satellite (TESS) continuously over 25 days in both Sector 1 (2019) and Sector 28 (2020).  Visual inspection of the TESS lightcurves shows that the Pre-search Data Conditioned Simple Aperture Photometry (PDCSAP) lightcurve appear to have additional trends added by the PDCSAP process, hence we adopt the Simple Aperture Photometry (SAP) lightcurve instead.  A Lomb-Scargle periodogram analysis of the SAP lightcurve from Sector 1 yields a broad double peaked periodogram with the peak covering periods from 17 to 23 days and highest individual peak powers at 17.8 and 22.7 days.  The Lomb-Scargle periodogram for Sector 28 has a power peak at the period of 18.1 days.  The TESS monitoring only covers $\sim$1 rotational period for this star, so does not have the time baseline necessary for a robust period determination.  Hence we adopt the 19.6$\pm$0.5 days period from our dedicated ground-based monitoring campaign.

With a comparatively long rotation period given its spectral type, HIP113201 is on the slow rotator branch. Using the \citet{Delorme2011} calibration with the 19.6$\pm$0.5 days period yields an age of 1.2$\pm$0.1 Gyr for this star.  The stellar properties of HIP 113201 are summarized in Table~\ref{tab:primaries}.

\subsection{HIP 36985/ GJ 282C }

With a separation of $>$50000 AU, HIP 36985 is one of the widest known companions, and is a triple component to the K dwarf binary GJ 282AB \citep{Poveda2009}.  \citet{Gaidos2014} fit the optical spectrum of HIP 36985 and find a best fit spectral type of M1, with best fit values of T$_\mathrm{eff}$ = 3744$\pm$82 K, luminosity = 0.044$\pm$0.01 L$_{\odot}$, radius = 0.51$\pm$0.05 R$_{\odot}$, and mass = 0.54$\pm$0.07 M$_{\odot}$. This M1 spectral type is independently confirmed by \citet{Alonso-Floriano2015}, from low-resolution optical spectroscopy with CAFOS at the 2.2 m Calar Alto telescope.  Using the $M_{K}$-mass relationship for stars with masses between 0.075 and 0.7 M$_{\odot}$ derived from orbital fits to 62 nearby binaries by \citet{Mann2019}, we estimate a mass of 0.57$\pm$0.01 M$_{\odot}$, matching well the mass estimate from spectroscopy. 

The X-ray emission of GJ282A and B \citep[individual~components~from~ROSAT~pointed~observations,][]{Schmitt2004} yield $log(L_{x}/L_{bol})$=-4.62 and -4.57, respectively.
These values are within the distribution of the Hyades members but slightly above the median values and well below the values observed
for Pleiades and AB Dor MG members (\citealt{Desidera2015}).
The log $R^{'}_{\rm HK}$ of GJ282A \citep[-4.39][]{Wright2004} is at the upper envelope of the Hyades members. The activity indicators then suggest an age slightly younger than the Hyades.

Lithium can be used as additional diagnostic: \citet{Ramirez2012} find
A(Li)=0.13$\pm$0.04 for HIP 36985, while lithium non-detection has been reported
by \citet{Mishenina2012} and \citet{Luck2017}. At the color of the star, a lithium non-detection implies an age older than about 550 Myr.
The spectroscopic analysis in the literature supports a metallicity close to solar ([Fe/H]=-0.09: \citet{Tabernero2017}; -0.12: \citet{Mishenina2012}; +0.01: \citet{Ramirez2012}; 0.00: \citet{Luck2017}; [M/H]=-0.05, \citet{Valenti2005}).


HIP 36985 has been monitored photometrically in $I$ and $Z$ by the All Sky Automated Survey \citep[ASAS,][]{Pojmanski2002}, with inhomogeneous sampling over a time baseline of $>$2000 days.  Using these data, \citet{Kiraga2012} and \citet{DiezAlonso2019} derive consistent rotation periods of 12.2$\pm$0.1 days and 12.16 days respectively.  HIP 36985 was also monitored by the Transiting Exoplanet Survey Satellite (TESS) continuously over 25 days.  Similarly as for the case of HIP 113201, visual inspection of the TESS lightcurves shows that the Pre-search Data Conditioned Simple Aperture Photometry (PDCSAP) lightcurve appear to have additional trends added by the PDCSAP process.  Thus, we again adopt the Simple Aperture Photometry (SAP) lightcurve instead.  A Lomb-Scargle periodogram analysis of the Simple SAP lightcurve from TESS yields a periodogram power peak at the period of $\sim$12.1 days.  


With a best rotational period of 12.2$\pm$0.1 days from literature values, 
HIP 36985 is a relatively fast rotator for its spectral type.
Since the best adopted period is shorter than the period at the convergence time on the slow rotator branch for this spectral type, gyrochronological relations such as those from \citet{Delorme2011} cannot be applied.  In this case, gyrochronology can only indicate that the age of the star is compatible with any age between 0 and the age of the Hyades.  Current best estimates for the age of the Hyades range from 625$\pm$50 Myr \citep{Delorme2011}, 650$\pm$70 Myr \citep{Martin2018}, and 750$\pm$100 Myr \citep{Brandt2015}.  Given that the rotation period of HIP 36985 is only slightly shorter than the period at the convergence time for an M1 spectral type, its age is likely similar to or slighter younger than the age of the Hyades, consistent with the age estimates from coronal and chromospheric emission.  

\begin{figure*}
\includegraphics[scale=1.2, angle=0, trim = 100 460 100 100]{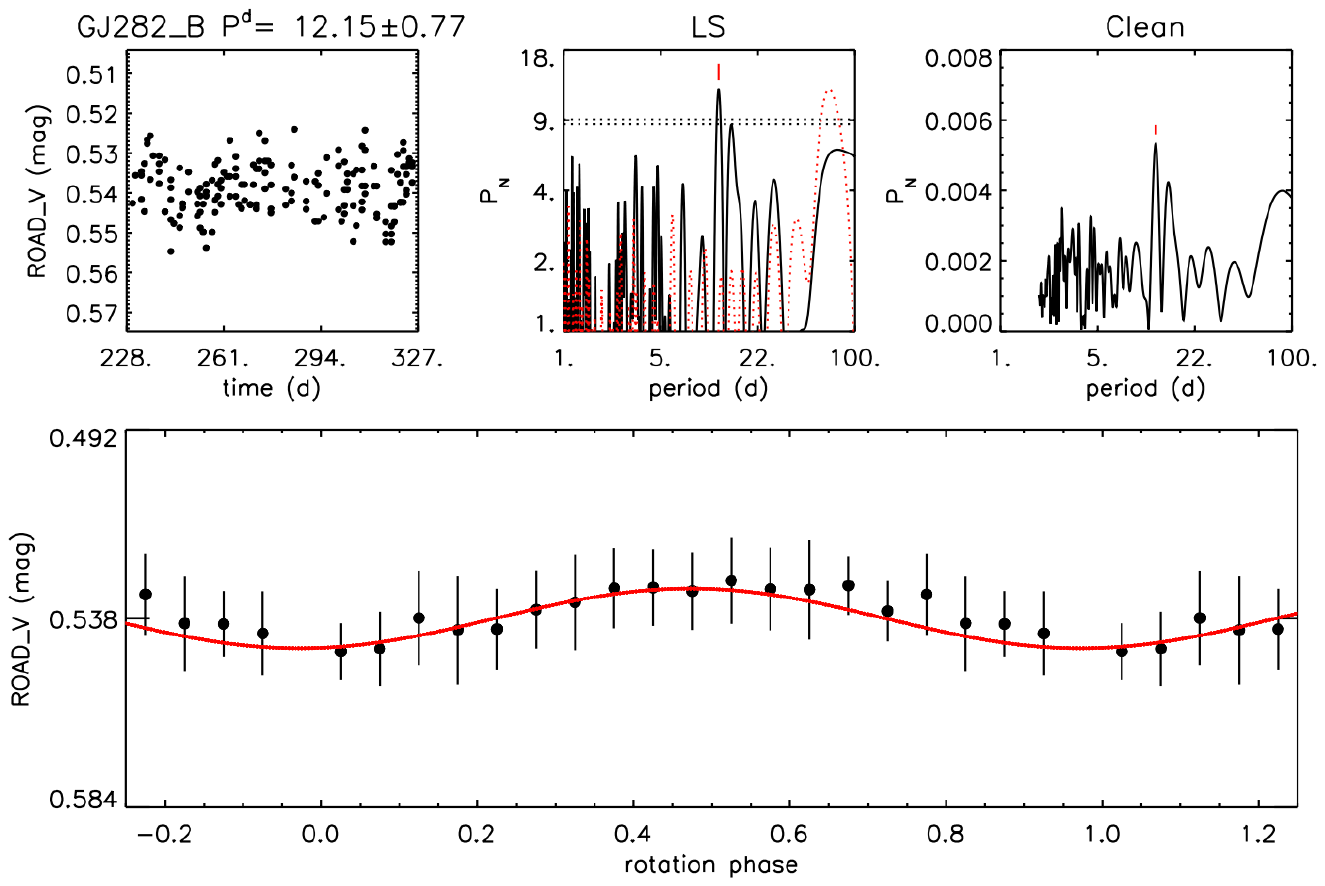}
\caption{Summary of periodogram analysis for GJ\,282B. Top-left panel: V-band magnitudes versus HJD (-2459000). Top-middle panel: Lomb-Scargle periodogram. The horizontal dotted lines show the 99.9\%  and 98\% confidence levels, whereas the red dotted line is the spectral window function. Top-right panel: the Clean periodogram. Bottom panel: the lightcurve phased with the 12.15-d rotation period. The red line is a sinusoidal fit to the phased light curve.}
\label{fig:periodogram_GJ282B}       
\end{figure*}

The two higher mass components of this system have K spectral types, thus are actually better targets for gyrochronological age dating, as they are older than the convergence time for these earlier spectral types.  Both were observed as part of ASAS, but no significant periodicities were found in these data. GJ 282A and B were covered in TESS Sector 7.  A LS periodogram analysis yielded the following periods for GJ 282 A and B: P$_A$= 13$\pm$3.5d and P$_B$= 14$\pm$4.0d, of insufficient precision for an accurate gyrochronological age dating.  Given the inconclusive results from the ASAS and TESS lightcurves, we conducted a photometric monitoring campaign for GJ 282A and B, from Jan 15 until April 20, 2021 for a total of 75 nights at the ROAD observatory (see Sect.\,2.1 for details). Observations were collected in the V and B filters and consisted of five consecutive frames per filter on each night, totaling 750 frames.  The magnitudes of GJ282A and B were extracted using aperture photometry together with two nearby stars that served as comparison (C; BD-03\,2003) and check (CK: HD\,61723) stars.  The comparison and check stars turned out to be constant in flux during the monitoring with a standard deviation $\sigma_V$ = 0.017 mag and $\sigma_B$ = 0.012 mag.  Both A and B components show a low level of variability with a standard deviation of their differential magnitude time series similar to that of the C$-$CK differential magnitudes with $\sigma_{V_A}$ = 0.017 mag; $\sigma_{B_A}$ = 0.013 mag and $\sigma_{V_B}$ = 0.015 mag; $\sigma_{B_B}$ = 0.015 mag. The periodogram analysis, carried out with Lomb-Scargle and CLEAN, did not reveal any significant periodicity for the A component. In contrast, the periodogram analysis of the B component showed a significant power peak in both V and B band at P = 12.10$\pm$0.77 days.  In Fig.~\ref{fig:periodogram_GJ282B}, we summarize the results of our periodogram analysis. To make the phase rotational modulation more easily visible, we averaged all magnitudes within bins of 0.05 in phase.

Using the gyrochronology relationship from \citet{Delorme2011} and adopting an age for the Hyades of 625 Myr, the period of 12.10$\pm$0.77 days for GJ 282B implies an age of 740$\pm$100 Myr.  This is compatible with the gyrochronological age limit of Hyades age or younger  that we found for HIP 36985, thus, combining as well with the age limits placed by lithium non-detection and activity indicators,  we adopt a system age similar to the Hyades for GJ 282AB / HIP 36985. 

The system has been considered for membership in the $\sim$300 Myr Ursa Major (UMa) moving group in several works \citep[e.g.~][]{Montes2001}.
Most recently, \citet{Baroch2021} recalculated the
galactocentric velocity for HIP 36985, finding 25.1, -2.5, and -7.8 km$~s^{-1}$
for the U, V, and W components, respectively, and suggest that it may be a member of the UMa moving group. These values are very similar to those of GJ 282A (U,V,W = 25.20$\pm$0.26, -3.07$\pm$0.25, -7.65$\pm$0.09 km$~s^{-1}$) listed by \citet{Tabernero2017}, further supporting the physical association of HIP 36985 with GJ 282AB.  \citet{Montes2001} find a mean velocity vector of the UMa nucleus of U, V, W = 14.9, 1.0, -10.7 km$~s^{-1}$; \citet{King2003} and \citet{Mamajek2010} find updated mean velocity vectors of U, V, W = 14.2$\pm$0.7, 2.8$\pm$1.3, -8.7$\pm$1.8  km$~s^{-1}$ and U, V, W = 15.0$\pm$0.4, 2.8$\pm$0.7, -8.1$\pm$1.0 km$~s^{-1}$ respectively. Both HIP 36985 and GJ 272A have velocity vectors close to but significantly outside the UMa nucleus.  Combined with the higher system age we find from gyrochronology, and consistent with estimates from coronal and chromospheric emission, we conclude that this stellar system is not part of the nucleus of the UMa group.
The stellar properties of HIP 36985 are summarized in Table~\ref{tab:primaries}.

\begin{center}
\begin{table*}
\caption{Properties of Primary Stars \label{tab:primaries}}
\begin{minipage}{160mm}
\begin{tabular}{ | c | c | c | c |}
\hline 
 & HIP 113201 & HIP 36985 & reference \\ \hline\hline
Spectral Type &  M1  &  M1 & \citet{Gaidos2014} \\ 
Effective Temperature & 3693$\pm$82 K & 3744$\pm$82 K & \citet{Gaidos2014} \\
Radius &  0.49$\pm$0.05 R$_{\odot}$ & 0.51$\pm$0.05 R$_{\odot}$ & \citet{Gaidos2014} \\
Mass (M$_{\odot}$) & 0.52$\pm$0.07 M$_{\odot}$ & 0.54$\pm$0.07 M$_{\odot}$ & \citet{Gaidos2014} \\
Mass (M$_{\odot}$) & 0.52$\pm$0.01 M$_{\odot}$ & 0.57$\pm$0.01 M$_{\odot}$ & using relationship from \citet{Mann2019} \\
Parallax (Gaia EDR3) & 42.49$\pm$0.22 mas &  70.27$\pm$0.13 mas & \citet{GaiaCollab2021} \\
Rotation period & 19.6$\pm$0.5 days & 12.2$\pm$0.1 days & this paper, \citet{Kiraga2012} \\
Gyrochronological Age & 1.2$\pm$0.1 Gyr & 740$\pm$100 Myr & this paper \\ \hline
\end{tabular}
\end{minipage}
\end{table*}
\end{center}


\section{Description of the parallel-tempering Markov chain Monte Carlo orbital fitting code \label{sec:mcmc}}

We used the parallel-tempering Markov chain Monte Carlo (PT-MCMC) ensemble sampler as implemented in \texttt{emcee v2.1.0} \citep{Foreman-Mackey2013} to fit the combination of radial velocity, direct imaging, and Gaia/Hipparcos astrometry available for each system.  The parallel-tempering MCMC method runs chains at different "temperatures" \citep{Earl2005}, where higher temperature chains sample large volumes of phase space, while lower temperature chains sample more precisely a local region of phase space. At the end of the MCMC run, only the results from the lowest temperature chain are retained, however, in the PT-MCMC implementation, the higher and lower temperature chains swap information, preventing the lower temperature chains from becoming trapped in local minima.  Thus, the PT-MCMC method is particularly useful for evaluating multimodal or complex posteriors.

\subsection{Simulating trial orbits\label{sec:dirvorbits}}
 
At each step of each MCMC chain, we simulate trial orbits using the following parameters (including the standard orbital elements):

\begin{itemize}
    \item $P$ -- orbital period (years)
    \item $i$ -- orbital inclination (degrees)
    \item $e$ -- orbital eccentricity
    \item $\gamma$ -- RV measurement offset (km~s$^{-1}$)
    \item $\Omega$ -- longitude of the ascending node (degrees)
    \item $\omega$ -- argument of periastron (degrees)
    \item $T_0$ -- time of periastron passage, in Julian days (JD)
    \item $d$ -- distance to system (pc)
    \item $M_{tot}$ -- total system mass (M$_{\odot}$)
    \item $M_2$ -- companion mass (M$_{\odot}$)
\end{itemize}

We construct simulated orbits following the approach given in \citet{Argyle2012}.
Time $t$ in julian date at each epoch is converted to mean anomaly $M$ as per:

\begin{equation}
M = \frac{2\pi}{P}(t - T_0)
\end{equation}

Mean anomaly $M$ is converted to eccentric anomaly $E$ by using the Newton-Raphson method to iteratively solve Kepler's equation:

\begin{equation}
M = E - sin E    
\end{equation}

Radial velocity variations were modeled at each epoch $t$ as:

\begin{equation}
    V(t) = K [cos(\omega) + \nu(t)) + e cos(\omega)]
\end{equation}

where $\omega$ is the argument of periastron, $\nu(t)$ is the true anomaly at epoch t, $e$ is the eccentricity, and the true anomaly $\nu(t)$ is related to the eccentric anomaly $E$ by:

\begin{equation}
    \nu = 2~arctan\left(\left(\sqrt{\frac{1+e}{1-e}}\right) tan\left(\frac{E}{2}\right)\right)
\end{equation}

As the arctan function in python only returns angles from $-\frac{\pi}{2}$ to $\frac{\pi}{2}$, we adopt the following form for $\nu$ to cover the full range of possible true anomaly values:

\begin{equation}
    \nu = 2~arctan2\left(\left(\sqrt{1+e}\right) sin\left(\frac{E}{2}\right), \left(\sqrt{1-e}\right) cos\left(\frac{E}{2}\right) \right)
\end{equation}

$K$ is the RV semi-amplitude, given by:

\begin{equation}
    K = \left(\frac{2 \pi G}{P(sec)}\right)^{\frac{1}{3}} \frac{M_2~sin(i)}{(1-e)^{2} M_{tot}^{\frac{2}{3}}}
\end{equation}

where $P(sec)$ is the orbital period in seconds, $M_2$ is the companion mass, $M_{tot}$ is the total system mass, $e$ is eccentricity and $i$ is the inclination of the system.  We adopt an additional offset parameter $\gamma$ for each RV data set, to take into account the systematic errors between instruments. 

Direct imaging relative companion positions (with respect to the stellar position) were modeled using the Thiele-Innes elements.
The semi-major axis $a$ of the system in AU is calculated from Kepler's 3rd law:

\begin{equation}
a = \left(P^2~M_{tot}\right)^{1/3}
\end{equation}

This is projected onto the sky using the Thiele-Innes elements:

\begin{align}
A &= a~(cos(\Omega)~cos(\omega) - sin(\Omega)~sin(\omega)~cos(i)) \\
B &= a~(sin(\Omega)~cos(\omega) + cos(\Omega)~sin(\omega)~cos(i)) \\
F &= a~(-cos(\Omega)~sin(\omega) - sin(\Omega)~cos(\omega)~cos(i)) \\
G &= a~(-sin(\Omega)~sin(\omega) + cos(\Omega)~cos(\omega)~cos(i))
\end{align}

At a given time $t$ / mean anomaly $M$ / eccentric anomaly $E$, then the projected separation (in AU) of the companion from the primary star in the x direction, y direction, or radially are given by:

\begin{align}
X &= cos(E) - e \\
Y &= \sqrt{1 - e^2}~sin(E) \\
R &= a~\sqrt{X^2 + Y^2}
\end{align}
    
Following a similar approach as \citet{Grandjean2019}, the tangential velocity of the companion at time $t$ is given by: 

\begin{align}
V_X &= - \frac{2~\pi~a^2~sin(E)}{P~R} \\
V_Y &= \frac{2~\pi~a^2~cos(E)}{P~R}~\sqrt{1-e^2}
\end{align}

Projecting onto the sky to obtain separations and velocities in right ascension and declination directions then provides the following companion positions and tangential velocities relative to the primary star:

\begin{align}
\Delta dec &= A~X + F~Y ~in~AU \\
\Delta ra &= B~X + G~Y  ~in~AU \\
\Delta v_{dec} &= \frac{A~V_X + F~V_Y}{a}~in~AU/yr \\
\Delta v_{ra} &= \frac{B~V_X + G~V_Y}{a}~in~AU/yr
\end{align}

To convert onto the on-sky projection requires a distance measurement; we include a fit to the Gaia DR2 distance in our likelihood function to help pin the distance of the system to an appropriate value.

We fit the radial velocity and direct imaging points directly; as epoch astrometry is not yet available for Gaia, we fit the average model tangential velocity in RA and Dec to the Hipparcos and Gaia proper motion anomaly measurements.  To correctly compare Gaia and Hipparcos proper motion measurements to our model orbits then requires also taking the (model-dependent) barycentric motion of the system into account.

\subsection{Simulating Hipparcos/Gaia proper motion anomalies}

For each primary star, the combination of Hipparcos and Gaia data yields three measurements: $\mu_H$, the proper motion as measured over the Hipparcos mission, $\mu_G$, the proper motion measured over the portion of the Gaia mission corresponding to a Gaia data release (e.g. Gaia DR2 or EDR3), and $\mu_{HG}$, the scaled positional difference between the Hipparcos and Gaia positional measurements over the full $\sim$25 year baseline between both missions \citep{Brandt2018, Brandt2019, Brandt2021}.  In the case of a single star, all three of these measurements will be the same.  A binary companion tugging on the primary star will cause deviations in proper motion on the timescale of the orbit of the companion, see e.g. Fig.~1 of \citet{Kervella2019}.  Thus, the "proper motion anomaly", the difference between the measured proper motion at the Hipparcos or Gaia epoch and the long-term, proper motion of the photocenter of the system, as defined in \citet{}{Kervella2019}:

\begin{equation}
\Delta \mu_{H/G2} = \mu_{H/G2} - \mu_{HG}, 
\label{eqn:pmanomaly}
\end{equation}

can be an important indicator of binarity and can provide additional orbital constraints. For both HIP 113201 and HIP 36985, the significant proper motion anomaly in both cases \citep{Kervella2019, Bonavita21} strongly pointed to the existence of the binary companion.  
For the purpose of orbit-fitting, the proper motion anomaly (an average quantity over a given mission timespan) is compared with either the instantaneous proper motion from the model at the midpoint of that mission timespan or the mean value of proper motion from the model over the mission timespan.  A shorter proper motion monitoring time thus will provide a proper motion anomaly measurement which is a better approximation to the instantaneous or mean model proper motion.  If a given mission timespan catches a part of the orbit with particularly high tangential velocity, a believable fit to the proper motion anomaly for that mission may not be possible.
Thus, in the future, for orbit-fitting, it would be preferable to directly fit epoch-by-epoch measurements.  However, epoch astrometry will not be available until Gaia DR4 and later.  For now, average measurements from astrometry over 1.5-2 year timespans, while not entirely accurate, can place valuable additional constraints on the orbit.  Because of this issue, we choose to fit Gaia DR2 proper motions instead of the more recent Gaia EDR3 proper motions, as Gaia DR2 covers a shorter on-sky period and hence provides a more "instantaneous" measurement.

The proper motion anomaly is measured relative to the photocenter of the system (assumed to be the primary star), whereas model quantities are calculated relative to the barycenter of the system.  Thus, to compare observed and model quantities, we must account for the barycenter position of the system relative to the photocenter. At a given time, $t$, the instantaneous on-sky tangential velocity (in mas) of the primary ($V_{primary}$) due to the pull of the secondary ($v_{secondary}$) relative to the barycenter of the system in right ascension and declination respectively is:

\begin{align}
V_{primary,~ra} &= -\frac{M_2}{M_{tot}}~v_{secondary,~ra} ~\frac{1000}{d}\\
V_{primary,~dec} &= -\frac{M_2}{M_{tot}}~v_{secondary,~dec} ~\frac{1000}{d}\\
\end{align}

However, as epoch astrometry is not available for intermediate Gaia data releases, what we measure instead is the average over all the Gaia or Hipparcos individual measurement epochs for each data release.  For each model orbit, we calculate this as the average of the tangential velocities for each of the observed epochs:

\begin{align}
\langle V_{primary,~ra} \rangle &= -\frac{M_2}{M_{tot}}~\langle v_{secondary,~ra}(t) \rangle ~ \frac{1000}{d}\\
\langle V_{primary,~dec} \rangle &= -\frac{M_2}{M_{tot}}~\langle v_{secondary,~dec}(t) \rangle ~ \frac{1000}{d}\\
\label{eqn:avgt}
\end{align}

For Hipparcos, we retrieved the observation dates using the Hipparcos intermediate observation app\footnote[1]{https://www.cosmos.esa.int/web/hipparcos/intermediate-data}
and for Gaia, we used the Gaia Observation forecast tool (GOST)\footnote[2]{https://gaia.esac.esa.int/gost/}to select observations within the Gaia DR2 observation period, calculated the instantaneous tangential velocity for each individual observation, and then adopted the average of these values to compare against the (averaged) Hipparcos or Gaia proper motion.

The tangential velocity calculated from the model gives the instantaneous motion on the sky due to the orbit of the secondary, whereas the proper motion anomaly calculated via Equation~\ref{eqn:pmanomaly} subtracts the long-term proper motion trend from the "instantaneous" Gaia or Hipparcos proper motion to approximate this same value.  However, all Gaia and Hipparcos proper motion values are calculated relative to the photocenter of the system (assumed to be the primary in this case, as the photometric contribution of the secondary is negligible), while the average model tangential velocity calculated via Equation~\ref{eqn:avgt} is relative to the barycenter of the system.  Thus, we must correct the proper motion anomaly measurement to the barycenter of the system.  This correction depends on the masses of both components of the binary, and hence, is model-dependent.  Again, following \citet{Grandjean2019}, the position of the star relative to the barycenter of the system due to reflex motion from the influence of the companion is given by:

\begin{align}
\Delta dec_* &= - \Delta dec~ \frac{M_2}{M_{tot}~d} \\
\Delta ra_* &= - \Delta ra~ \frac{M_2}{M_{tot}~d}\\    
\end{align}

The barycenter proper motion, as calculated from the scaled Hipparcos-Gaia positional difference is then given by:

\begin{align}
\mu_{bary~ra} &= -\frac{M_2}{M_{tot}~d} \frac{\langle \Delta ra_* \rangle_G - \langle \Delta ra_* \rangle_H}{\langle t_H \rangle - \langle t_G \rangle} \\
\mu_{bary~dec} &= -\frac{M_2}{M_{tot}~d} \frac{\langle \Delta dec_* \rangle_G - \langle \Delta dec_* \rangle_H}{\langle t_H \rangle - \langle t_G \rangle}
\end{align}

To compare between model and data quantities, we then subtract out the model barycenter proper motion from the observed proper motion anomaly (in RA and DEC respectively):

\begin{align}
\Delta \mu_{bary} = \mu_{H/G2} - \mu_{HG} - \mu_{bary}
\end{align}

We can then compare $\Delta \mu_{ref}$ in RA and DEC with $V_{primary,~ra}$ and $V_{primary,~dec}$. Errors for $\Delta \mu_{ref}$ are added in quadrature.

\subsection{Likelihood function and priors}

Combining direct imaging, radial velocity, and Hipparcos / Gaia astrometry, our final combined likelihood function for the PT-MCMC runs is:

\begin{multline}
\mathcal{L} = -\frac{1}{2} \Bigg[ \sum_{i=1}^{N_{DI}} \left(\frac{x_i^{DI} - f_i^{DI}}{\sigma_i^{DI}}\right)^{2} + \sum_{i=1}^{N_{DI}} \left(\frac{y_i^{DI} - f_i^{DI}}{\sigma_i^{DI}}\right)^{2}  + \\  
\sum_{i=1}^{N_{inst}} \sum_{j=1}^{N_{RV}} \left(\frac{RV_j - (f_j^{RV} + \gamma_i)}{\sigma_j^{RV}}\right)^{2}  + 
\left(\frac{\Delta \mu_{bary,H,RA} - V_{primary,H,RA}}{\sigma_{\Delta \mu_{bary,H,RA}}}\right)^{2} +  \\ 
\left(\frac{\Delta \mu_{bary,H,DEC} - V_{primary,H,DEC}}{\sigma_{\Delta \mu_{bary,HIP,DEC}}}\right)^{2}  +   
\left(\frac{\Delta \mu_{bary,G,RA} - V_{primary,G,RA}}{\sigma_{\Delta \mu_{bary,G,RA}}} \right)^{2}  + \\
\left(\frac{\Delta \mu_{bary,G,DEC} - V_{primary,G,DEC}}{\sigma_{\Delta \mu_{bary,G,DEC}}}\right)^{2} + 
\left(\frac{\varpi_{DR2} - (1000/d)}{\sigma_{\varpi_{DR2}}}\right)^{2} \Bigg]
\end{multline}

where $x_i^{DI}$ and $y_i^{DI}$ are the direct imaging offsets of the secondary from the primary position in RA and Declination ($\Delta ra$ and $\Delta dec$ from Section~\ref{sec:dirvorbits}) respectively for each data epoch, $f_i^{DI}$ are the model predictions for the secondary position relative to the primary at each data epoch, $RV_{j}$ is the measured radial velocity at each data epoch, $f_j^{RV}$ is the model prediction for radial velocity at that data epoch, $\gamma_i$ are the instrumental offsets between the model radial velocity and the measured radial velocity for each instrument $i$, $\Delta \mu_{bary}$ values are the barycenter corrected proper motion anomaly in RA and Dec for Hipparcos and Gaia respectively, $\varpi_{DR2}$ is the measured Gaia DR2 parallax in mas, and $(1000/d)$ is the model prediction for the parallax of the system (with distance $d$ given in pc).  The values labeled as $\sigma$ are the respective errors on each measured property.  For properties that combine multiple measurements (e.g. various $\Delta \mu_{bary}$ values), $\sigma_{\Delta \mu_{bary}}$ has been calculated as the sum in quadrature of all the constituent measured values.  The PT-MCMC fits do not actually serve to put useful constraints on the distance to the system; the fit to the Gaia DR2 parallax is included rather as an additional prior, to weight the PT-MCMC runs to distance values which are consistent with the measured Gaia parallax.

For all PT-MCMC runs, we adopted the following uniform priors:
\begin{itemize}
    \item $P$ -- 0 to 70 years
    \item $i$,$\Omega$,$\omega$ -- 0 to 2$\pi$ radians
    \item $e$ -- 0 to 1
    \item $\gamma$ -- -20 km s$^{-1}$ to 20 km s$^{-1}$, fit separately for each instrument used 
    \item $T_0$ -- uniform between  2450000 days < $T_{0}$ < 2464500 days for HIP 113201 and between 2454000 days < $T_0$ < 2466000 days for HIP 36985
    \item $M_{tot}$ -- 0-2 M$_{\odot}$
    \item $M_2$ -- 0-0.5 M$_{\odot}$
\end{itemize}

\section{Orbital Fits\label{sec:orbits}}

\subsection{HIP 113201}

For direct imaging points, we fit the VLT-SPHERE astrometric data points presented in \citet{Bonavita21}.  HIP 113201 also has multiple years (spanning 2008 to 2017) of radial-velocity monitoring using the HARPS instrument on the 3.6 m telescope at the La Silla observatory.
We derived radial velocity measurements from all spectra available on the ESO archive using the SAFIR (Spectroscopic data via Analysis of the Fourier Interspectrum Radial velocities) pipeline described in \citet{Galland_SAFIR}.  These measurements are presented in Appendix~\ref{app:harps113201}.  

We performed PT-MCMC fits to 1) the combination of direct imaging and RV data by itself, 2) the combination of direct imaging, RV, and both HIP/Gaia proper motion anomalies, 3) the combination of direct imaging, RV, and only the HIP proper motion anomaly (still using Gaia astrometry to correct for barycentric motion and fit for the system distance) and 4) the combination of direct imaging, RV, and only the Gaia proper motion anomaly.  For most runs, we ran the PT-MCMC sampler described in Section~\ref{sec:mcmc} with 15 temperatures, 50 walkers and 40000 steps, and used the final 20000 steps of the coldest temperature walkers for figures shown here.  For the fit to just the direct imaging and RV points, the chains did not reach convergence within 40000 steps, so we re-ran this fit with 100 walkers.  For all MCMC runs, we inspected the walkers from the coldest temperature chain by eye to determine when the chain had converged.  

\begin{figure*}
\includegraphics[width=\textwidth]{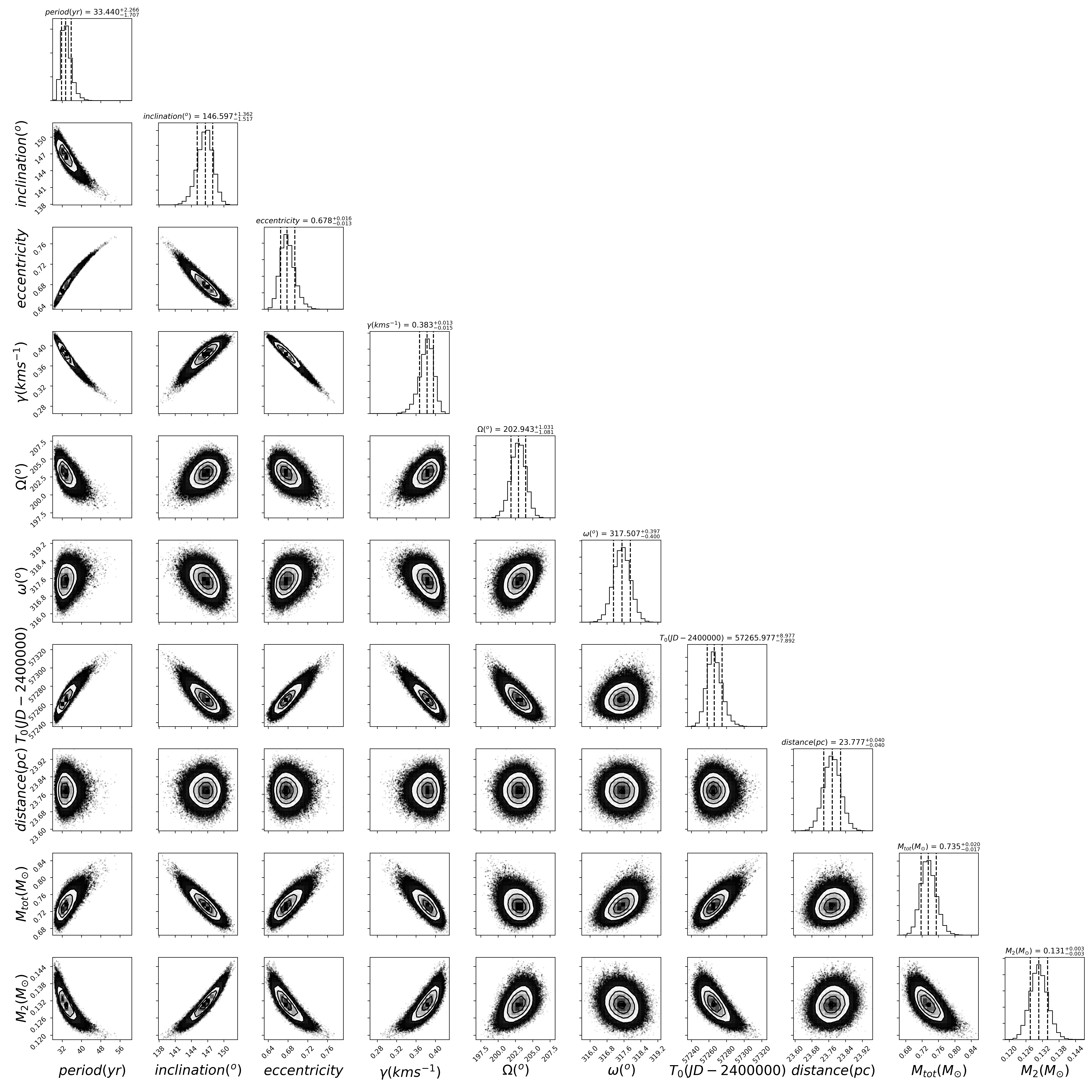}
\caption{The corner plot for PT-MCMC fit to the orbit of HIP 113201AB, incorporating direct imaging, radial velocity and the Hipparcos proper motion anomaly in the orbital fit.
}
\label{fig:fit_corner_HIP113201}       
\end{figure*}

\begin{figure*}
\includegraphics[width=0.55\textwidth]{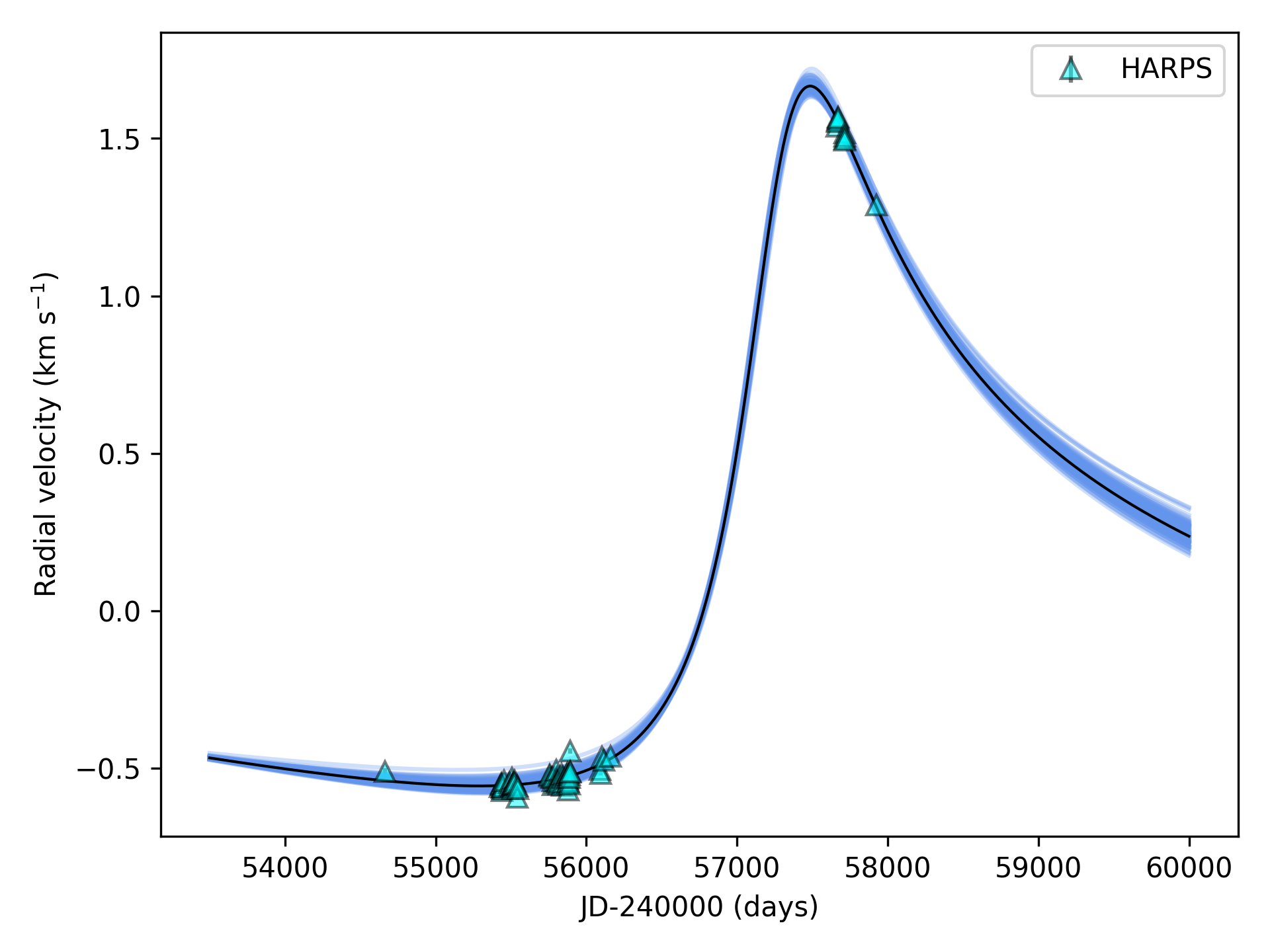}
\includegraphics[width=0.45\textwidth]{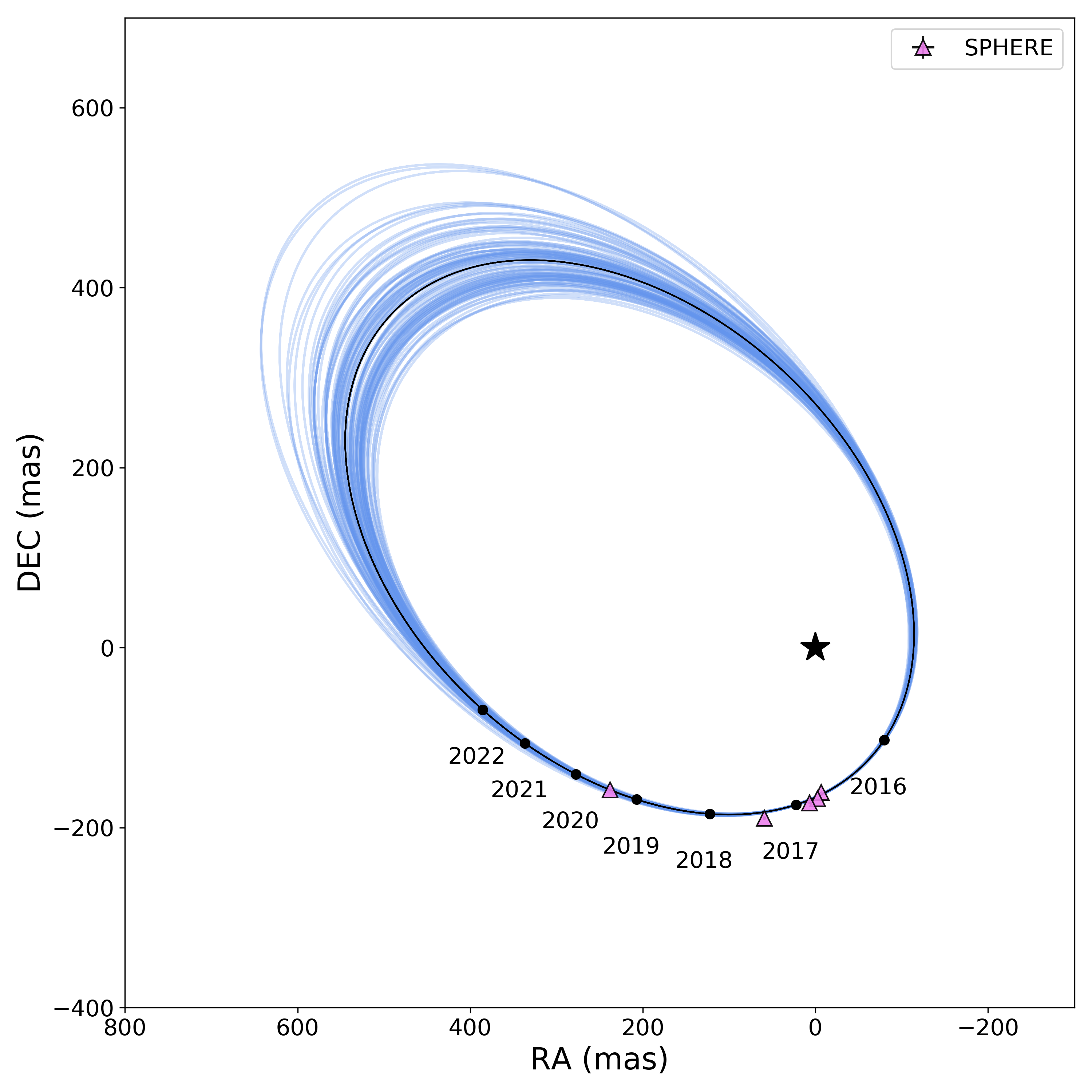}
\caption{\textit{Left:} Radial velocity data vs. model comparison for HIP 113201AB.  HARPS data points are plotted as cyan triangles.  The best fit orbit from the direct imaging, radial velocity and the Hipparcos proper motion anomaly PT-MCMC run is plotted as a solid black line; blue lines depict 100 random orbits taken from the final converged PT-MCMC posterior pdf. \textit{Right:} Direct imaging data vs. model comparison for HIP 113201AB.  SPHERE data points are plotted as lavender triangles.  The best fit orbit from the direct imaging, radial velocity and the Hipparcos proper motion anomaly PT-MCMC run is plotted as a solid black line; blue lines depict 100 random orbits taken from the final converged PT-MCMC posterior pdf.  The position of the primary is depicted with a black star symbol.  
}
\label{fig:RV_HIP113201}       
\end{figure*}

Best parameters derived from each MCMC run are presented in Table~\ref{tab:HIP113201fits}.  
We adopt the DI+RV+Hipparcos fit as the best overall fit to the system.  Corner plots for all parameters of this fit are presented in Fig.~\ref{fig:fit_corner_HIP113201}.  The best fit orbit, as well as 100 orbits randomly selected from the posterior probability distribution function (henceforth pdf) are plotted alongside the RV and direct imaging data in Fig.~\ref{fig:RV_HIP113201}.  The best fit orbit and 100 randomly selected orbits from the posterior of the model tangential motion on the sky compared to the barycenter-corrected Hipparcos and DR2 proper motion anomalies are presented in Fig.~\ref{fig:fit_tangential_HIP113201}.  Similar plots for the other fits are presented in Appendix~\ref{app:HIP113201_dirv} for the fit to just direct imaging and RV data, Appendix~\ref{app:HIP113201_dirvhipgaia} for the fit to direct imaging data, RV data, and both Gaia and Hipparcos proper motion anomalies, and Appendix~\ref{app:HIP113201_dirvgaia} for the fit to direct imaging data, RV data, and only the Gaia proper motion anomaly.  Unfortunately, the Gaia DR2 epoch caught the orbit in a period with extreme tangential motion -- the averaging technique we used here to compare the tangential motion of the companion to the Gaia/HIP astrometry is not appropriate for this epoch, given the extreme motion on the sky, especially in right ascension.  Forcing a fit of the high-precision Gaia measurement to the uncertain model astrometry over the full 1.5 year Gaia DR2 observation period drives the period to larger values.  Consequently, to compensate for the longer period from the Gaia constraint, the fit to direct imaging, RV, and Gaia/HIP astrometry produces an anomalously high mass for the primary.  This dynamical mass is incompatible with both spectroscopic and photometric mass estimates for this star \citep{Gaidos2014, Bonavita21, Mann2019} -- with an M1 spectral type, we do not expect a mass beyond 0.6 M$_{\odot}$ for HIP 113201A.  Hence, we adopt the fit using only the Hipparcos constraint for our final values, as the Hipparcos epochs of observations cover a much more gradual, well-behaved portion of the orbital motion.

\begin{table*}
\caption{HIP 113201 orbital fits \label{tab:HIP113201fits}}
\begin{tabular}{ c c c c c}
Parameter & DI+RV & \textbf{DI+RV+HIP} & DI+RV+Gaia & DI+RV+HIP+Gaia \\ \hline\hline
P (years)  &  $27.773_{-1.787}^{+2.290}$ &  $33.440_{-1.707}^{+2.266}$ & $38.568_{-1.794}^{+1.861}$ &  $38.412_{-1.754}^{+1.811}$ \\
$T_0$ (BJD) & $2457239.591_{-11.053}^{+12.328}$ &  $2457265.977_{-7.892}^{+8.977}$ & $2457278.773_{-8.073}^{+7.398}$ &  $2457278.316_{-7.880}^{+7.331}$ \\
e & $0.628_{-0.020}^{+0.022}$ &  $0.678_{-0.013}^{+0.016}$ & $0.741_{-0.009}^{+0.009}$ & $0.740_{-0.009}^{+0.008}$ \\
i (deg) & $151.893_{-2.547}^{+2.554}$ & $146.597_{-1.517}^{+1.362}$ & $137.783_{-0.442}^{+0.434}$ &  $137.810_{-0.437}^{+0.434}$ \\
$\omega$ (deg) & $317.130_{-0.425}^{+0.426}$ & $317.507_{-0.400}^{+0.397}$ & $318.814_{-0.389}^{+0.386}$ & $318.811_{-0.383}^{+0.383}$ \\
$\Omega$ (deg) & $205.018_{-1.276}^{+1.212}$ & $202.943_{-1.081}^{+1.031}$ & $207.939_{-1.163}^{+1.241}$ &  $207.982_{-1.161}^{+1.211}$ \\
$\gamma$ (km s$^{-1}$) & $0.431_{-0.021}^{+0.019}$ &  $0.383_{-0.015}^{+0.013}$ &  $0.346_{-0.009}^{+0.009}$ & $0.347_{-0.009}^{+0.009}$ \\
$M_{tot}~(M_{\odot})$& $0.685_{-0.021}^{+0.023}$ &  $0.735_{-0.017}^{+0.020}$ & $0.935_{-0.013}^{+0.013}$ & $0.934_{-0.013}^{+0.013}$ \\
$M_2~(M_{\odot})$ & $0.146_{-0.008}^{+0.010}$ & $0.131_{-0.003}^{+0.003}$ & $0.125_{-0.001}^{+0.001}$ &  $0.125_{-0.001}^{+0.001}$ \\
$M_1~(M_{\odot})$ & $0.539_{-0.030}^{+0.031}$ & $0.604_{-0.020}^{+0.022}$ & $0.809_{-0.013}^{+0.013}$ & $0.809_{-0.013}^{+0.013}$ \\
\hline\hline
\end{tabular}
\end{table*}

\begin{figure*}
\includegraphics[scale=0.5]{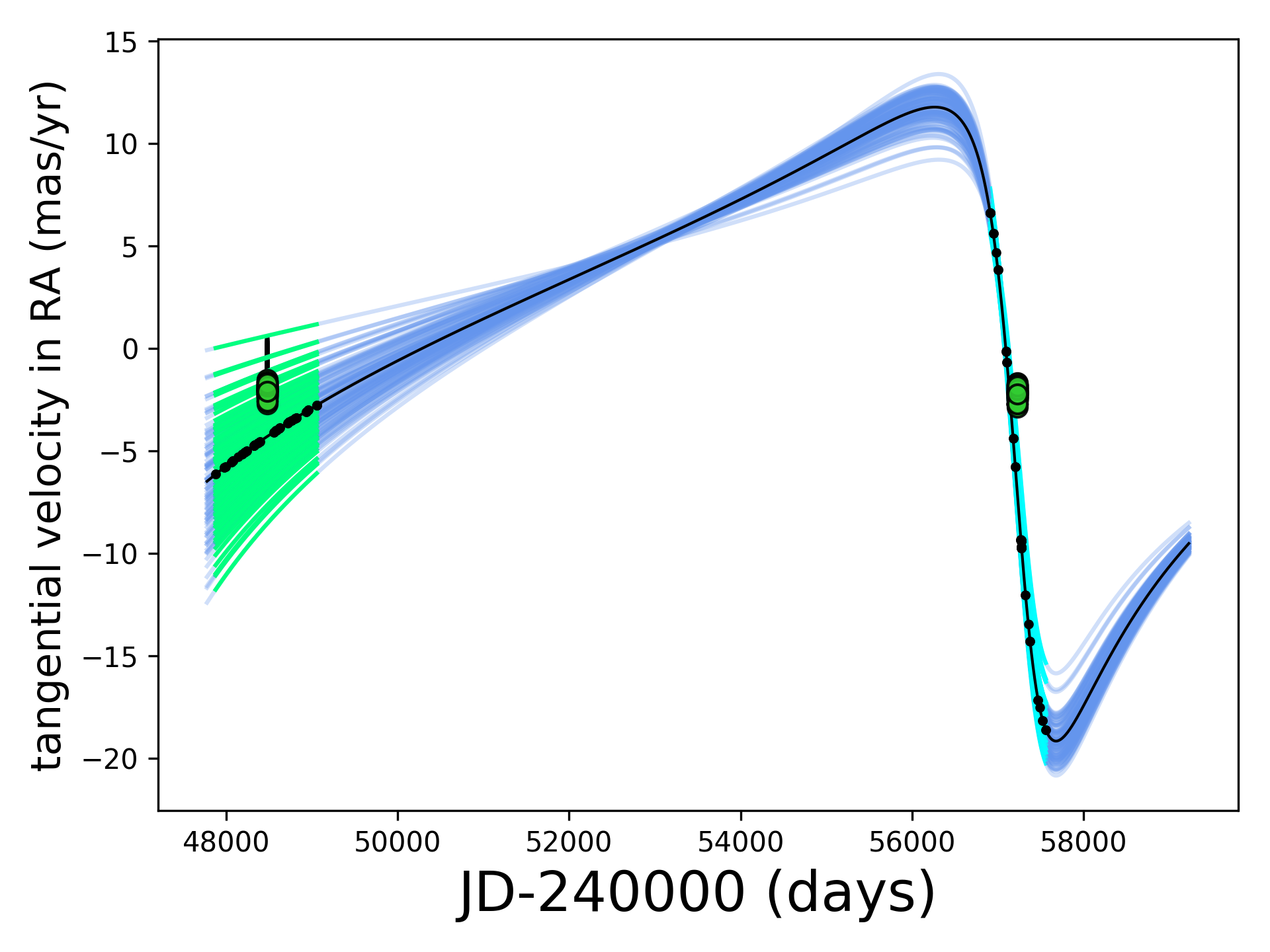}
\includegraphics[scale=0.5]{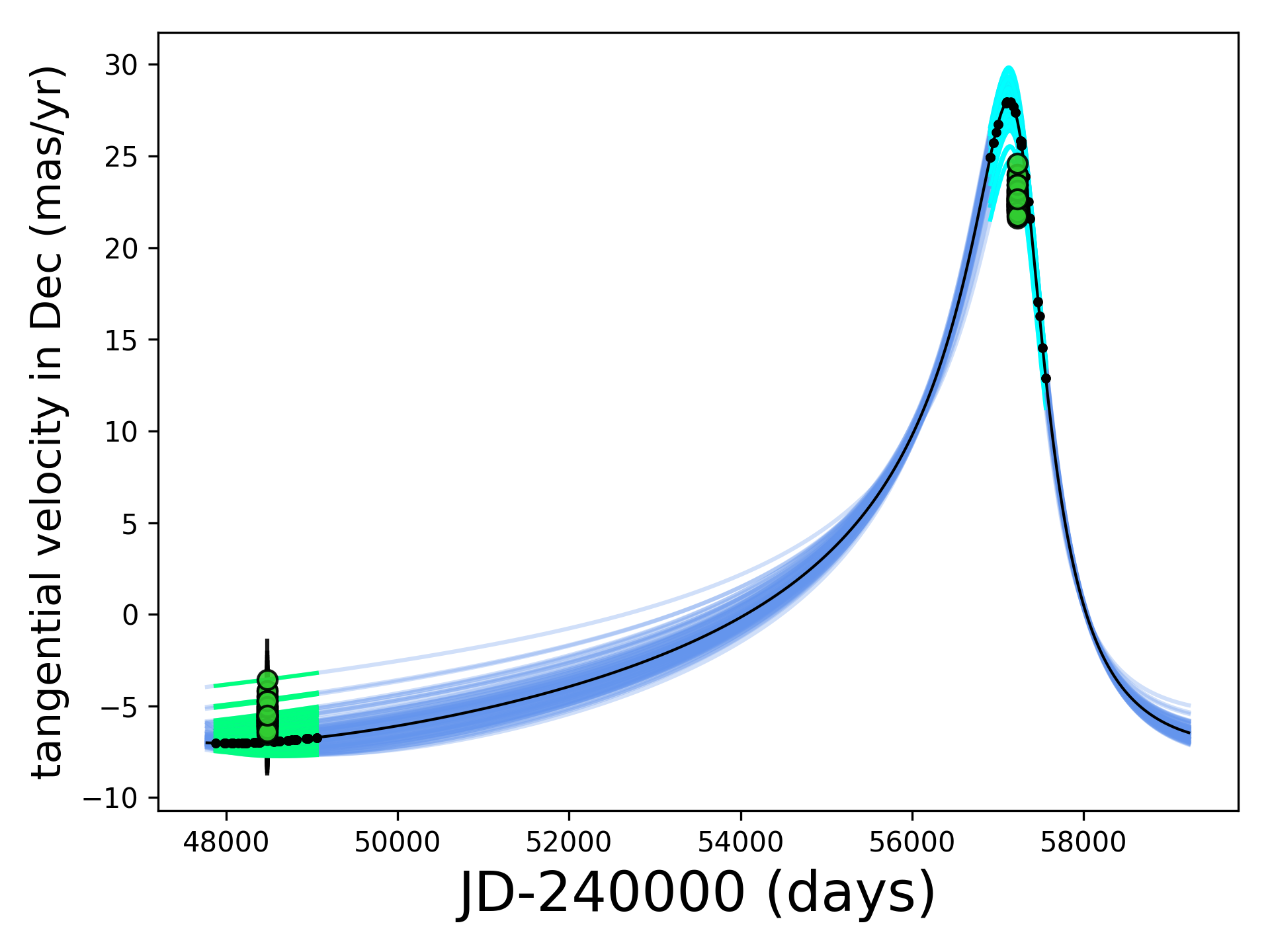}
\caption{Model tangential velocity compared to Hipparcos and Gaia barycenter-corrected proper motion anomalies for HIP 113201AB. The tangential velocity in RA is plotted in the left panel and the tangential velocity in Declination is plotted in the right panel.  The best fit orbit from the direct imaging, radial velocity and the Hipparcos proper motion anomaly PT-MCMC run is plotted as a solid black line; blue lines depict 100 random orbits taken from the final converged PT-MCMC posterior pdf.  Hipparcos and Gaia barycenter-corrected proper motion anomalies are plotted as green points for the same 100 random orbits; because the barycenter correction depends on primary and secondary mass, these vary slightly depending on the orbit selected.  The Hipparcos mission lifetime for the 100 random orbits is highlighted in green; the Gaia DR2 observation period is highlighted in cyan.  The small circle points depict the dates at which Hipparcos and Gaia measurements were acquired.  
}
\label{fig:fit_tangential_HIP113201}       
\end{figure*}

\subsection{HIP 36985}

HIP 36985 has extensive multi-instrument monitoring available for both direct imaging and radial velocity observations.  For our fits, we included the 3 VLT-NaCo direct imaging observations presented in Table 3 of \citet{Baroch2021} and the VLT-SPHERE astrometric data points presented in Table 7 of \citet{Bonavita21}.  HIP 36985 has multiple years (spanning 2014 to 2019) of radial-velocity monitoring using the HARPS instrument on the 3.6 m telescope at the La Silla observatory \citep{Grandjean2020} and a significant proper motion anomaly between Hipparcos and Gaia DR2 measurements \citep{Kervella2019}.  We again draw astrometry values from the Hipparcos-Gaia Catalog of Accelerations \citep{Brandt2018, Brandt2021} to  account for systematics on the sky and ensure all astrometry is in the Gaia DR2 reference frame. Additional radial velocity monitoring with FEROS and CARMENES is presented in \citet{Baroch2021}.   We fit all HARPS, FEROS and CARMENES data points, taken from Table A.1 of \citet{Baroch2021}.  To account for the different RV offsets, we fit 3 offset terms ($\gamma_1$, $\gamma_2$, $\gamma_3$), one per instrument.  

We performed two PT-MCMC fits to combinations of these data: 1) only direct imaging and RV points and 2) direct imaging, RV, and the HIP/Gaia proper motion anomalies.  In both cases, we ran the PT-MCMC sampler for 15 temperatures, 50 walkers and 50000 steps, and used the final 20000 steps of the coldest temperature walker for figures shown here.  We inspected 100 walkers from the coldest temperature chain by eye to determine when the chain had converged.  The Gaia DR2 observations for HIP 36985 cover a less extreme part of the orbital motion than for HIP 113201, thus, we found that including both Hipparcos and Gaia proper motion anomalies in our fits improved the overall fit and produced physically consistent masses for both primary and secondary.     

Best parameters derived from each PT-MCMC run, as well as from the fit from \citet{Baroch2021}, are presented in Table~\ref{tab:HIP36985fits}.  We find comparable fits from fitting only the direct imaging and the radial velocity data and from the fit to all direct imaging data, RV data, and Hipparcos/Gaia proper motion anomalies.  Thus, we adopt the DI+RV+Hipparcos/Gaia fit as the best overall fit to the system.  This fit matches well with the model parameters found using a similar method and the \texttt{orvara} fitting code from \citet{Baroch2021}.  This is unsurprising, as we fit nearly the same data points as \citet{Baroch2021}, with a very similar method.  Our fit additionally incorporates 3 SPHERE-SHINE observations, providing slightly stronger constraints on the orbital motion.

\begin{figure*}
\includegraphics[width=\textwidth]{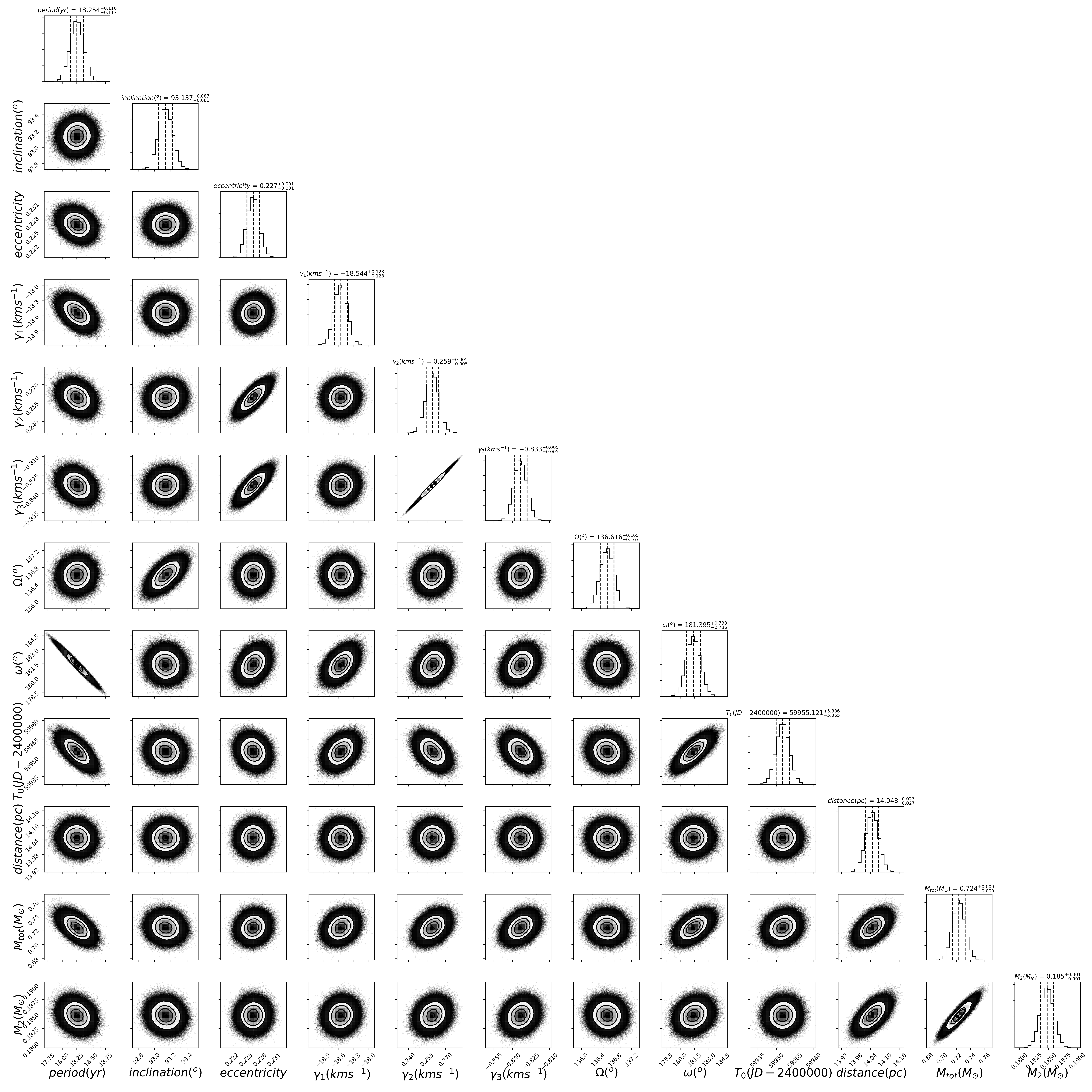}
\caption{The corner plot for PT-MCMC fit to the orbit of HIP 36985AB, incorporating direct imaging, radial velocity and both Hipparcos and Gaia proper motion anomalies in the orbital fit.
}
\label{fig:fit_corner_HIP36985}       
\end{figure*}

\begin{figure*}
\includegraphics[width=0.55\textwidth]{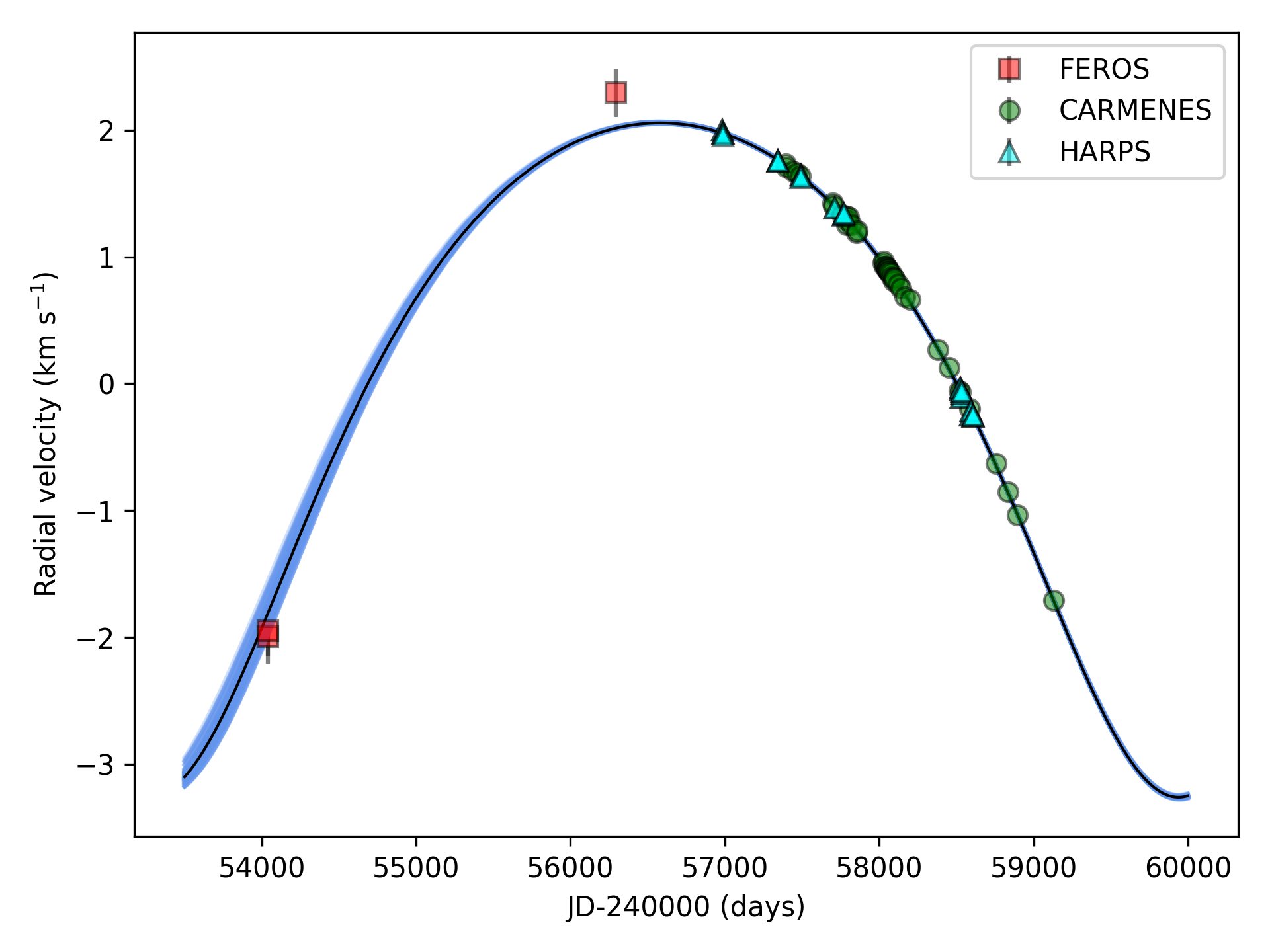}
\includegraphics[width=0.45\textwidth]{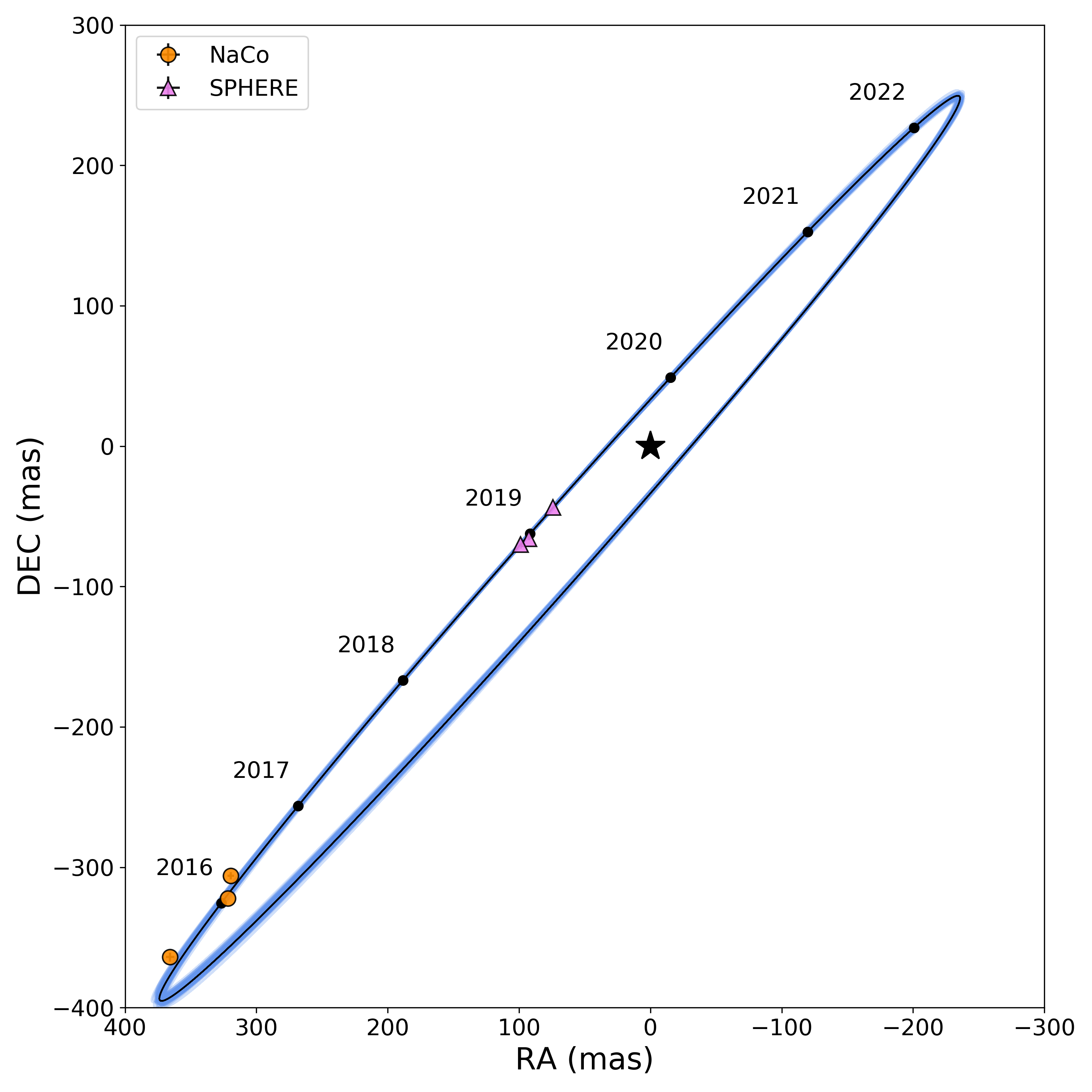}
\caption{\textit{Left:} Radial velocity data vs. model comparison for HIP 36985AB.  FEROS data points are plotted as red squares, CARMENES data points are plotted as green circles, and HARPS data points are plotted as cyan triangles.  The best fit orbit to the direct imaging data, radial velocity data and both the Hipparcos and Gaia proper motion anomalies is plotted as a solid black line; blue lines depict 100 random orbits taken from the final converged PT-MCMC posterior pdf. \textit{Right:} Direct imaging data vs. model comparison for HIP 36985AB.  NaCo data points are plotted as orange circles and SPHERE data points are plotted as lavender triangles.  The best fit orbit to the direct imaging data, radial velocity data and both the Hipparcos and Gaia proper motion anomalies is plotted as a solid black line; blue lines depict 100 random orbits taken from the final converged PT-MCMC posterior pdf.  The position of the primary is depicted with a black star symbol.  
}
\label{fig:RV_HIP36985}       
\end{figure*}

\begin{figure*}
\includegraphics[scale=0.5]{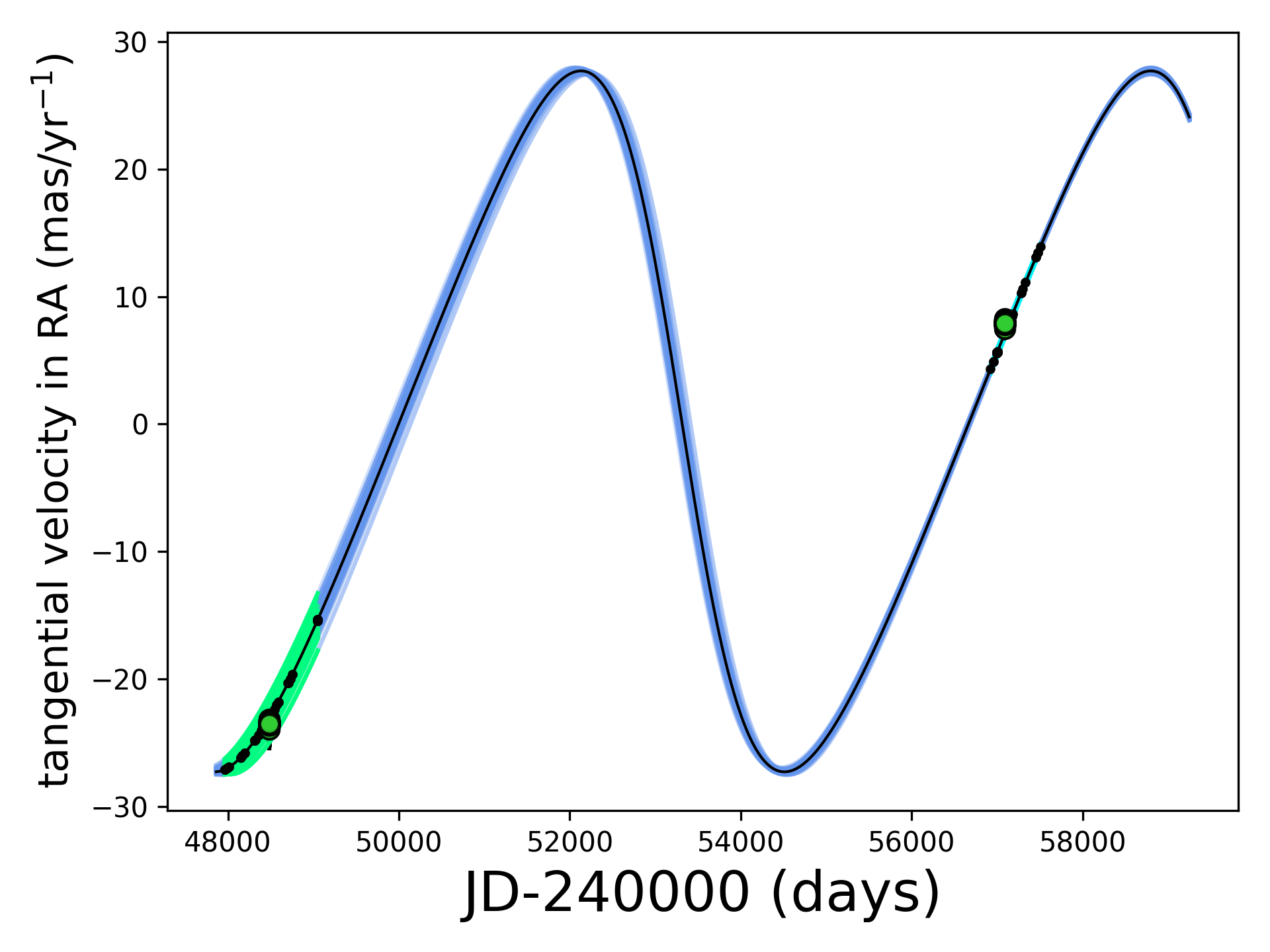}
\includegraphics[scale=0.5]{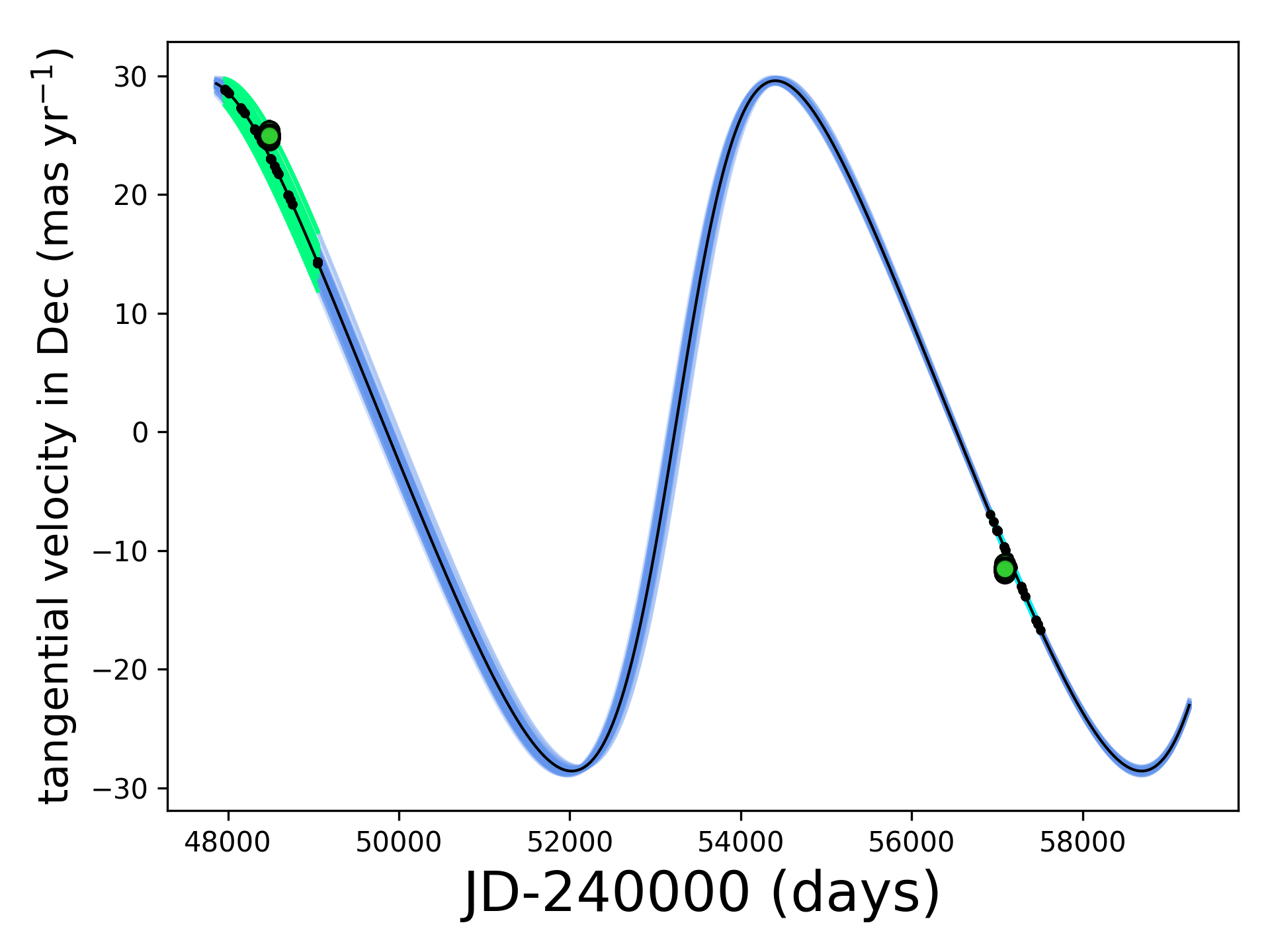}
\caption{Model tangential velocity compared to Hipparcos and Gaia barycenter-corrected proper motion anomalies for HIP 36985AB.  The tangential velocity in RA is plotted in the left panel and the tangential velocity in Declination is plotted in the right panel. The best fit orbit from the direct imaging, radial velocity and the Hipparcos proper motion anomaly PT-MCMC run is plotted as a solid black line; blue lines depict 100 random orbits taken from the final converged PT-MCMC posterior pdf.  Hipparcos and Gaia barycenter-corrected proper motion anomalies are plotted as green points for the same 100 random orbits; because the barycenter correction depends on primary and secondary mass, these vary slightly depending on the orbit selected.  The Hipparcos mission lifetime for the 100 random orbits is highlighted in green; the Gaia DR2 observation period is highlighted in cyan.  The small circle points depict the dates at which Hipparcos and Gaia measurements were acquired.  
}
\label{fig:fit_tangential_HIP36985}       
\end{figure*}

Corner plots for all parameters of the DI+RV+Hipparcos/Gaia fit are presented in Fig.~\ref{fig:fit_corner_HIP36985}.  The best fit orbit, as well as 100 orbits randomly selected from the posterior pdf are plotted alongside the RV and direct imaging data in Fig.~\ref{fig:RV_HIP36985}.  The best fit orbit and 100 randomly selected orbits from the posterior of the model tangential motion on the sky compared to the barycenter-corrected Hipparcos and DR2 proper motion anomalies are presented in Fig.~\ref{fig:fit_tangential_HIP36985}.  

\begin{table*}
\caption{HIP 36985 orbital fits\label{tab:HIP36985fits}.}
\begin{tabular}{ c c c c }
Parameter & DI+RV & {\bf DI+RV+HIP+Gaia} & \citet{Baroch2021}$^{1}$ \\ \hline\hline
P (years)  &  $19.942_{-0.326}^{+0.348}$ & $18.254_{-0.117}^{+0.116}$ & $18.045^{+0.372}_{-0.485}$\\
$T_0$ (BJD) & $2459890.444_{-15.525}^{+14.494}$ & $2459955.121_{-5.365}^{+5.336}$ & $2460004_{-91}^{+90}$ \\
e & $0.225_{-0.001}^{+0.002}$ & $0.227_{-0.001}^{+0.001}$ & $0.213_{-0.010}^{+0.010}$ \\ 
i (deg) & $93.070_{-0.088}^{+0.088}$ & $93.137_{-0.086}^{+0.087}$ & $93.96_{-0.55}^{+0.55}$\\
$\omega$ (deg) & $171.225_{-1.938}^{+1.875}$ & $181.395_{-0.736}^{+0.738}$ & $183.0_{-1.4}^{+1.5}$\\
$\Omega$ (deg) & $136.472_{-0.168}^{+0.169}$ & $136.616_{-0.167}^{+0.165}$ & $136.95_{-0.46}^{+0.46}$\\
$\gamma_1$ (FEROS, km s$^{-1}$) & $-19.267_{-0.164}^{+0.165}$ & $-18.544_{-0.128}^{+0.128}$ &  $-18.35_{-0.12}^{+0.11}$ ($\gamma$) \\
$\gamma_2$ (HARPS, km s$^{-1}$) & $0.242_{-0.006}^{+0.006}$ & $0.259_{-0.005}^{+0.005}$ & $-0.20_{-0.12}^{+0.11}$ ($\gamma_H$) \\
$\gamma_3$ (CARMENES, km s$^{-1}$) & $-0.852_{-0.006}^{+0.006}$ & $-0.833_{-0.005}^{+0.005}$ &  0.89$_{-0.12}^{+0.11}$ ($\gamma_C$) \\
$M_{tot}~(M_{\odot})$& $0.695_{-0.012}^{+0.012}$ & $0.724_{-0.009}^{+0.009}$ &   \\
$M_2~(M_{\odot})$ &  $0.187_{-0.001}^{+0.001}$ & $0.185_{-0.001}^{+0.001}$ & $0.1881_{-0.0047}^{+0.0048}$\\
$M_1~(M_{\odot})$ & $0.508_{-0.012}^{+0.011}$ & $0.539_{-0.007}^{+0.008}$ & $0.554^{+0.058}_{-0.049}$ \\
\hline\hline
\end{tabular}
\newline
\newline
1: \citet{Baroch2021} define radial velocity offsets $\gamma$ relative to the FEROS observations, while we define a separate offset for each instrument.
\end{table*}

\section{Discussion -- Comparison of companion dynamical masses with existing mass estimates \label{sec:discussion}}

Before full orbital fitting was possible for these objects, a number of authors estimated the mass of all components of these systems via model isochrones.  Our dynamical mass measurements thus provide valuable tests both to photometric estimates of very low-mass star masses as well as to empirical mass-luminosity relationships from orbital fitting.  Here we compare our dynamical masses to both model estimates based on photometry and other dynamical mass measurements in the literature.

\citet{Bonavita21} estimate component masses using BT-Settl pre-MS isochrones \citep{Allard2014} with SPHERE photometry and Gaia EDR3 distances.  Their age estimates for HIP 113201 and HIP 36985 were derived from a mix of kinematic and gyrochronological age analysis.  Adopting ages of 750$\pm$250 Myr and 500$\pm$200 Myr respectively, \citet{Bonavita21} find primary and secondary masses of 0.53 M$_{\odot}$ and 0.10 M$_{\odot}$ for HIP 113201AB, and 0.58 M$_{\odot}$ and 0.19 M$_{\odot}$ for HIP 36985AB.  

\citet{Mann2019} derive an empirical M$_{K}$-Mass relation for stellar masses between 0.075 and 0.7 M$_{\odot}$ by finding total system masses from 62 nearby M-dwarf binaries.  No K-band photometric measurements are available for HIP 36985B and HIP 113201B, but using the reddest value $\Delta$mag measurements from \citet{Bonavita21} and assuming that the companions are uniformly redder than their primaries (i.e. that $\Delta$mag(K) will be less than the $\Delta$mag value we adopt from bluer SPHERE measurements), we can estimate minimum masses using the M$_{K}$-mass expression from \citet{Mann2019}.  For HIP 113201B, \citet{Bonavita21} report 5 $\Delta$J measurements at different epochs ranging from 3.54 to 4.09 mag.  We adopt the median value of these, and the standard deviation as the error: $\Delta$J = 3.93$\pm$0.31, yielding a minimum mass using the \citet{Mann2019} relationship of 0.099$\pm$0.010 M$_{\odot}$.
Adopting $\Delta H_3$=2.848$\pm$0.245 for HIP 36985B from \citet{Bonavita21}, we estimate a minimum mass using the \citet{Mann2019} relationship of 0.174$\pm$0.021 M$_{\odot}$.

Combining existing HARPS radial velocity measurements with SPHERE astrometry and the Hipparcos/Gaia proper motion anomaly measurement, \citet{Bonavita21} find a dynamical mass for HIP 36985B of 0.180$\pm$0.007 M$_{\odot}$.  Using the \texttt{orvara} orbit-fitting code \citep{Brandt2021b}, \citet{Baroch2021} fit HARPS, CARMENES, and FEROS radial velocity measurements, VLT-NaCo direct imaging measurements and the Hipparcos/Gaia proper motion anomaly, finding dynamical masses of 0.554$^{+0.058}_{-0.049}$ M$_{\odot}$ and 0.1881$ ^{+0.0048}_{-0.0047}$ M$_{\odot}$ for the primary and secondary respectively.   

Our best fit dynamical masses for both primaries agree well with photometric estimates, the empirical M$_{K}$-Mass relation from \citet{Mann2019}, and also existing dynamical mass measurements in the literature.  Both primaries are young main-sequence stars, thus, their model masses derived from their luminosity should be insensitive to the age of the star.

\begin{figure}
\includegraphics[width=\columnwidth]{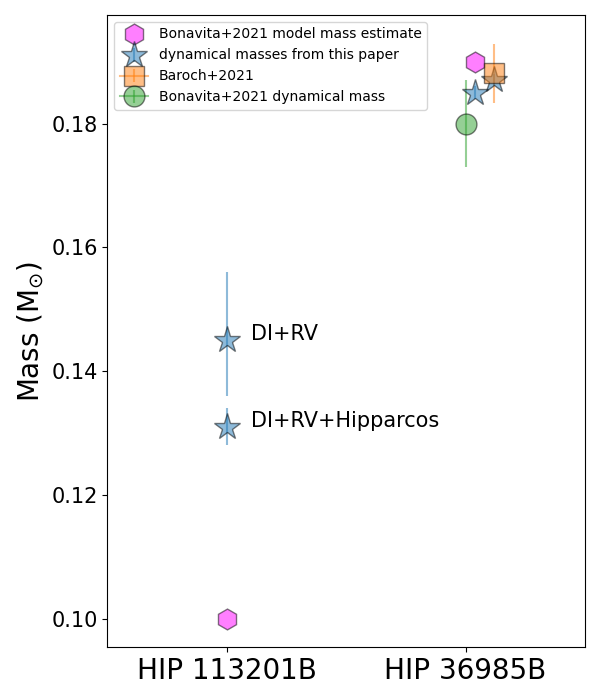}
\caption{Comparison of companion dynamical masses from this paper, dynamical mass from \citet{Baroch2021}, and model estimate and dynamical masses from \citet{Bonavita21}.  Mass values for HIP 36985B have been slightly offset from each other to improve readability.  While all dynamical mass measurements and mass estimates agree well for HIP 36985B, the dynamical mass measurements and model mass estimate diverge considerably for HIP 113201B, which is extremely faint given the observed dynamical effects it produces.
}
\label{fig:companion_mass_comparison}       
\end{figure}

\begin{figure}
\includegraphics[width=\columnwidth]{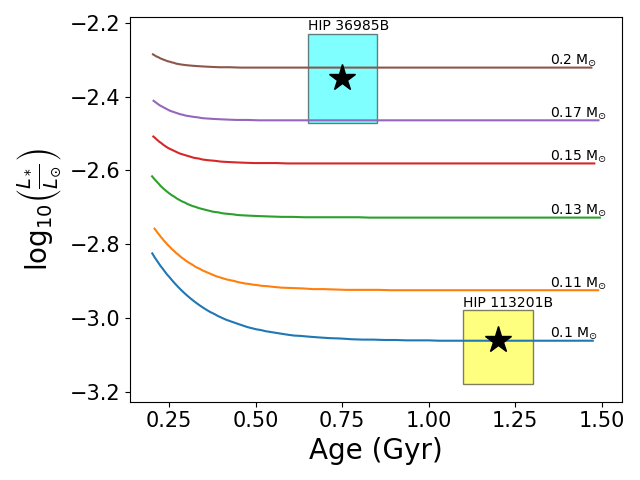}
\caption{Best age and $L_{bol}$ ranges for HIP 113201B (yellow rectangle) and HIP 36985B (cyan rectangle) overplotted on model evolutionary tracks for very low mass stars from \citet{Baraffe2015}.  Both companions have reached the main sequence, hence model mass estimates for these objects should be age independent for system ages greater than $\sim$0.6 Gyr.
}
\label{fig:BCAH98_model_predictions}       
\end{figure}

A comparison of dynamical masses and model mass estimates for HIP 113201B and HIP 36985B is presented in Fig.~\ref{fig:companion_mass_comparison}.  All mass estimates and measurements for HIP 36985B agree on a mass of around 0.19 $M_{\odot}$.  There is considerably more divergence between mass estimates for the lower mass HIP 113201B companion.  Estimates combining photometry and the system age with model isochrones find a mass of $\sim$0.1 M$_{\odot}$; we find a considerably higher dynamical mass of 0.13-0.15 M$_{\odot}$. We only find a lower limit to mass with the relationship from \citet{Mann2019}, but this is consistent with our dynamical mass estimate.
However, the 750$\pm$250 Myr age adopted for HIP 113201B in \citet{Bonavita21} is considerably lower than the age of 1.2$\pm$0.1 Gyr adopted here. To evaluate whether this is the source of the divergence between the dynamical mass and the model mass estimate from \citet{Bonavita21}, in Fig.~\ref{fig:BCAH98_model_predictions}, we overplot luminosity as a function of age using the \citet{Baraffe2015} models for objects with masses between 0.1 and 0.25 M$_{\odot}$ with the bolometric luminosities and age ranges of the two companions considered in this paper.  We estimated bolometric luminosities for HIP 113201B and HIP 36985B using the J-band bolometric correction from \citet{Filippazzo2015}, with the 2MASS $J$ measurements for each primary stars and the $\Delta(mag)$ values in the SPHERE $J2$ band reported in \citet{Bonavita21}.  We adopted a mid-M spectral type for each companion and used the M6 J-band bolometric correction from \citet{Filippazzo2015}, as this was the earliest spectral type covered in this paper.  The bolometric corrections from \citet{Filippazzo2015} do not change much between M6-M8 spectral types and we expect a roughly similar bolometric correction value for M4-M5 spectral types.  At the younger ages reported for both stars in \citet{Bonavita21} and the older age ranges adopted herein, both companions have clearly reached the main sequence, thus the mass estimate based on luminosity will be largely age independent.  We then firmly conclude that while all mass estimates for HIP 36985B agree well, HIP 113201B is anomalously faint given its observed dynamical impacts on its system.  An undetected brown dwarf companion to HIP 113201B could be a natural explanation for this apparent discrepancy.
Adding the luminosity from a 30 M$_{Jup}$ brown dwarf to the luminosity of HIP 113201B as modeled by the 0.1 M$_{\odot}$ model track from \citet{Baraffe2015} would make almost no difference to the total luminosity.  From the \citet{Baraffe2015} models, the $\Delta$magnitude in H band at 1 Gyr between a 0.1 M$_{\odot}$ star and a 30 M$_{Jup}$ companion would be 5.47 mag, so the total system brightness would change by less than 1\% in $H$-band. Meanwhile, the presence of such an unseen companion would bring the total system mass up to 0.13 $M_{\odot}$, in line with the dynamical data. Such a companion at a close separation could easily elude detection in all of the existing data.

\section{Conclusions\label{sec:conclusions}}

We present updated ages, orbital fits, and dynamical masses for HIP 113201AB and HIP 36985AB, two M1 + mid-M dwarf binary systems monitored as part of the SPHERE SHINE survey \citep{Desidera2021, Langlois2021, Vigan2020}.  The main results of this work are as follows:

\begin{itemize}
    \item To robustly determine the age of both systems via gyrochronology, we undertook a photometric monitoring campaign for HIP 113201 and for GJ 282AB, the two wide K star companions to HIP 36985, using the 40 cm telescope Remote Observatory Atacama Desert (ROAD) located in the
Atacama Desert.  HIP 113201 is a slow rotator with a period of 19.6$\pm$0.5 days, corresponding to a gyrochronological age of 1.2$\pm$0.1 Gyr using the relationship from \citet{Delorme2011}.  
HIP 36985 is a wide companion to the K star binary system GJ 282AB \citep{Poveda2009}.  The M1 star HIP 36985 has a rotation period shorter than the period at the convergence time on the slow rotator branch for this spectral type, so only an upper limit on age for this star can be placed from gyrochronological relations such as those from \citet{Delorme2011}.  However, its K star companions have reached convergence, hence we use the rotational periods for the K star components of the system to set the gyrochronological age of the system as a whole. Using the gyrochronology relationship from \citet{Delorme2011} and adopting an age for the Hyades of 625 Myr, the period of 12.10$\pm$0.77 days for GJ 282B implies an age of 750$\pm$100 Myr.  This is compatible with the gyrochronological age limit of 650 Myr that we found for HIP 36985, thus, combining as well with the age limits placed by lithium non-detection and activity indicators,  we adopt a system age similar to the Hyades for GJ 282AB / HIP 36985.  These system ages are sufficiently old that we expect that all components of these binaries will have reached the main sequence. 
    \item To derive dynamical masses for all components of the HIP 113201AB and HIP 36985AB systems, we used the parallel-tempering Markov chain Monte Carlo sampler implemented in \texttt{EMCEE} \citep{Foreman-Mackey2013} to fit a combination of radial velocity, direct imaging, and Gaia and Hipparcos astrometry.
    \item As epoch astrometry is not available yet for Gaia, we did not fit individual Hipparcos and Gaia data points, but rather, the proper motion anomaly \citep{Kervella2019} at both the Gaia and Hipparcos epochs. 
    \item For HIP 113201AB, combining direct imaging, radial velocity, and both Hipparcos and Gaia proper motion anomalies yield an unphysically large primary mass of $\sim$0.8 M$_{\odot}$ given the M1 spectral type of HIP 113201A, as well as a considerably longer period compared to fits of just the direct imaging and RV data. The proper motion anomaly method averages over 1.5-2 years of measurement -- we found in the case of HIP 113201AB that the Gaia proper motion anomaly did not serve as an appropriate constraint due to the extreme orbital motion of the companion in RA during the Gaia DR2 observation epochs.  Fitting the Hipparcos proper motion anomaly (which still relies on Gaia DR2 measurements to correct the barycenter motion) alongside the direct imaging and RV data yields a primary mass of $\sim$0.6 M$_{\odot}$, marginally consistent with the M1 spectral type of the primary, and a secondary mass of 0.13 M$_{\odot}$.  Fitting only the direct imaging and radial velocity data yields a primary mass of 0.54$\pm$0.03 M$_{\odot}$, fully consistent with its M1 spectral type, and a secondary mass of $\sim$0.145 M$_{\odot}$.  The secondary masses derived with and without including Hipparcos/Gaia data are all considerably more massive than the 0.1 M$_{\odot}$ estimated mass from the photometry of the companion \citep{Bonavita21}.  Thus, the dynamical impacts of this companion suggest that there is more mass in the system than expected from its photometry.  An undetected brown dwarf companion to HIP 113201B could be a natural explanation for this apparent discrepancy.  At an age $>$1 Gyr, a 30 M$_{Jup}$ companion to HIP 113201B would make a negligible ($<$1$\%$) contribution to the system luminosity, but could have strong dynamical impacts. 
    \item For HIP 36985AB, the dynamical masses found in this work for both the primary and secondary agree well with the photometric estimates of component masses, the masses estimated from the \citet{Mann2019} $M_{K}$-- mass relationship, and previous dynamical masses in the literature \citet{Baroch2021}.
\end{itemize}

In the end, the two systems studied here are rare and not exemplary of M dwarf multiple systems.  HIP 36985AB + GJ 282AB is an uncommon K+M dwarf quaternary system. In this paper, we have shown preliminary evidence that HIP 113201AB may comprise of an equally rare M dwarf star + brown dwarf triple system.  Both systems present interesting challenges to formation mechanisms for low-mass stellar multiple systems.

\begin{acknowledgements}

B.B acknowledges funding by the UK Science and Technology Facilities Council (STFC) grant no. ST/M001229/1.  B.B. would like to acknowledge valuable conversations with Trent Dupuy and Mariangela Bonavita which improved this manuscript.  This work has been supported by the PRIN-INAF 2019 "Planetary systems at young ages (PLATEA)" and the ASI-INAF agreement n.2018-16-HH.0.  A.Z. acknowledges support from the FONDECYT Iniciaci\'on en investigaci\'on project number 11190837.  SPHERE is an instrument designed and built by a consortium
consisting of IPAG (Grenoble, France), MPIA (Heidelberg, Germany), LAM
(Marseille, France), LESIA (Paris, France), Laboratoire Lagrange
(Nice, France), INAF–Osservatorio di Padova (Italy), Observatoire de
Genève (Switzerland), ETH Zurich (Switzerland), NOVA (Netherlands),
ONERA (France) and ASTRON (Netherlands) in collaboration with
ESO. SPHERE was funded by ESO, with additional contributions from CNRS
(France), MPIA (Germany), INAF (Italy), FINES (Switzerland) and NOVA
(Netherlands).  SPHERE also received funding from the European
Commission Sixth and Seventh Framework Programmes as part of the
Optical Infrared Coordination Network for Astronomy (OPTICON) under
grant number RII3-Ct-2004-001566 for FP6 (2004–2008), grant number
226604 for FP7 (2009–2012) and grant number 312430 for FP7
(2013–2016). We also acknowledge financial support from the Programme National de
Plan\'etologie (PNP) and the Programme National de Physique Stellaire
(PNPS) of CNRS-INSU in France. This work has also been supported by a grant from
the French Labex OSUG@2020 (Investissements d’avenir – ANR10 LABX56).
The project is supported by CNRS, by the Agence Nationale de la
Recherche (ANR-14-CE33-0018). It has also been carried out within the frame of the National Centre for Competence in 
Research PlanetS supported by the Swiss National Science Foundation (SNSF). MRM, HMS, and SD are pleased 
to acknowledge this financial support of the SNSF. Finally, this work has made use of the the SPHERE
Data Centre, jointly operated by OSUG/IPAG (Grenoble),
PYTHEAS/LAM/CESAM (Marseille), OCA/Lagrange (Nice), Observatoire de
Paris/LESIA (Paris), and Observatoire de Lyon, also supported by a grant from Labex 
OSUG@2020 (Investissements d’avenir – ANR10 LABX56). We thank P. Delorme and E. Lagadec (SPHERE Data
Centre) for their efficient help during the data reduction
process. 

\end{acknowledgements}

\bibliographystyle{aa}
\bibliography{main}

\begin{appendix}
\onecolumn
\section{HARPS RV measurements for HIP 113201 \label{app:harps113201}}

\begin{longtable}{ccc}
\caption{HARPS RV measurements for HIP 113201\label{tab:harps113201}} \\
\hline
BJD & RV ($km~s^{-1}$) & $\sigma_{RV}$ \\
\hline
\endfirsthead
\caption{continued} \\
\hline 
BJD & RV ($km~s^{-1}$) & $\sigma_{RV}$ \\
\hline
\endhead
\hline
\endfoot
\hline
\endlastfoot
2454664.942628 & -0.126406 & 0.009298 \\
2455425.617006 & -0.176033 & 0.007683 \\
2455435.695372 & -0.169375 & 0.005741 \\
2455436.701546 & -0.16539 & 0.004445 \\
2455438.813786 & -0.188146 & 0.00451 \\
2455439.815963 & -0.177052 & 0.00633 \\
2455444.740529 & -0.182145 & 0.005319 \\
2455446.700449 & -0.174255 & 0.005969 \\
2455453.726252 & -0.156608 & 0.005474 \\
2455488.638743 & -0.163103 & 0.004592 \\
2455494.646924 & -0.170389 & 0.005191 \\
2455505.541255 & -0.148593 & 0.00565 \\
2455507.545121 & -0.173555 & 0.004206 \\
2455510.524249 & -0.182202 & 0.005202 \\
2455512.531042 & -0.169319 & 0.005213 \\
2455515.53953 & -0.16941 & 0.005647 \\
2455517.555014 & -0.164319 & 0.005198 \\
2455522.569904 & -0.156187 & 0.006022 \\
2455524.572185 & -0.165444 & 0.007246 \\
2455542.539473 & -0.173678 & 0.005523 \\
2455543.535958 & -0.208701 & 0.008258 \\
2455544.558634 & -0.181086 & 0.005247 \\
2455755.921135 & -0.141719 & 0.007519 \\
2455756.894528 & -0.137068 & 0.007811 \\
2455770.899546 & -0.145386 & 0.005521 \\
2455776.799114 & -0.168146 & 0.006999 \\
2455777.778547 & -0.157029 & 0.005795 \\
2455779.767168 & -0.14573 & 0.007128 \\
2455802.853552 & -0.121752 & 0.007428 \\
2455804.793732 & -0.146126 & 0.008218 \\
2455805.688388 & -0.15717 & 0.00733 \\
2455809.742199 & -0.1597 & 0.00502 \\
2455809.742199 & -0.160686 & 0.005017 \\
2455816.712072 & -0.165838 & 0.005475 \\
2455836.701836 & -0.135323 & 0.004697 \\
2455839.715079 & -0.170729 & 0.007837 \\
2455842.591598 & -0.146251 & 0.006444 \\
2455845.66473 & -0.15985 & 0.006158 \\
2455871.555095 & -0.13781 & 0.004719 \\
2455873.579215 & -0.149079 & 0.00553 \\
2455874.589856 & -0.16367 & 0.007273 \\
2455875.552353 & -0.131093 & 0.005773 \\
2455878.579752 & -0.184755 & 0.008848 \\
2455879.564412 & -0.159054 & 0.008728 \\
2455880.545148 & -0.145538 & 0.005418 \\
2455888.585156 & -0.137011 & 0.005362 \\
2455889.599282 & -0.167447 & 0.006516 \\
2455890.595921 & -0.144902 & 0.004693 \\
2455892.595738 & -0.061266 & 0.008363 \\
2455893.564938 & -0.126182 & 0.005129 \\
2455894.575535 & -0.129591 & 0.00521 \\
2456088.860071 & -0.120262 & 0.008209 \\
2456096.892691 & -0.13366 & 0.006298 \\
2456102.944659 & -0.081753 & 0.005298 \\
2456115.875017 & -0.091465 & 0.006876 \\
2456161.76294 & -0.077645 & 0.006753 \\
2457662.708925 & 1.922066 & 0.006259 \\
2457666.556159 & 1.940951 & 0.004246 \\
2457666.56985 & 1.93638 & 0.004319 \\
2457669.611427 & 1.951621 & 0.0039 \\
2457669.625813 & 1.947327 & 0.00367 \\
2457712.549426 & 1.901318 & 0.006183 \\
2457712.563812 & 1.881134 & 0.006266 \\
2457713.558946 & 1.890442 & 0.005669 \\
2457713.569616 & 1.878793 & 0.006548 \\
2457923.852302 & 1.671995 & 0.006387 \\ \hline
\end{longtable}


\vspace{5cm}

\FloatBarrier

\section{HIP 113201 orbital fits -- direct imaging and RV only fit \label{app:HIP113201_dirv}}

\begin{figure*}
\includegraphics[width=\textwidth]{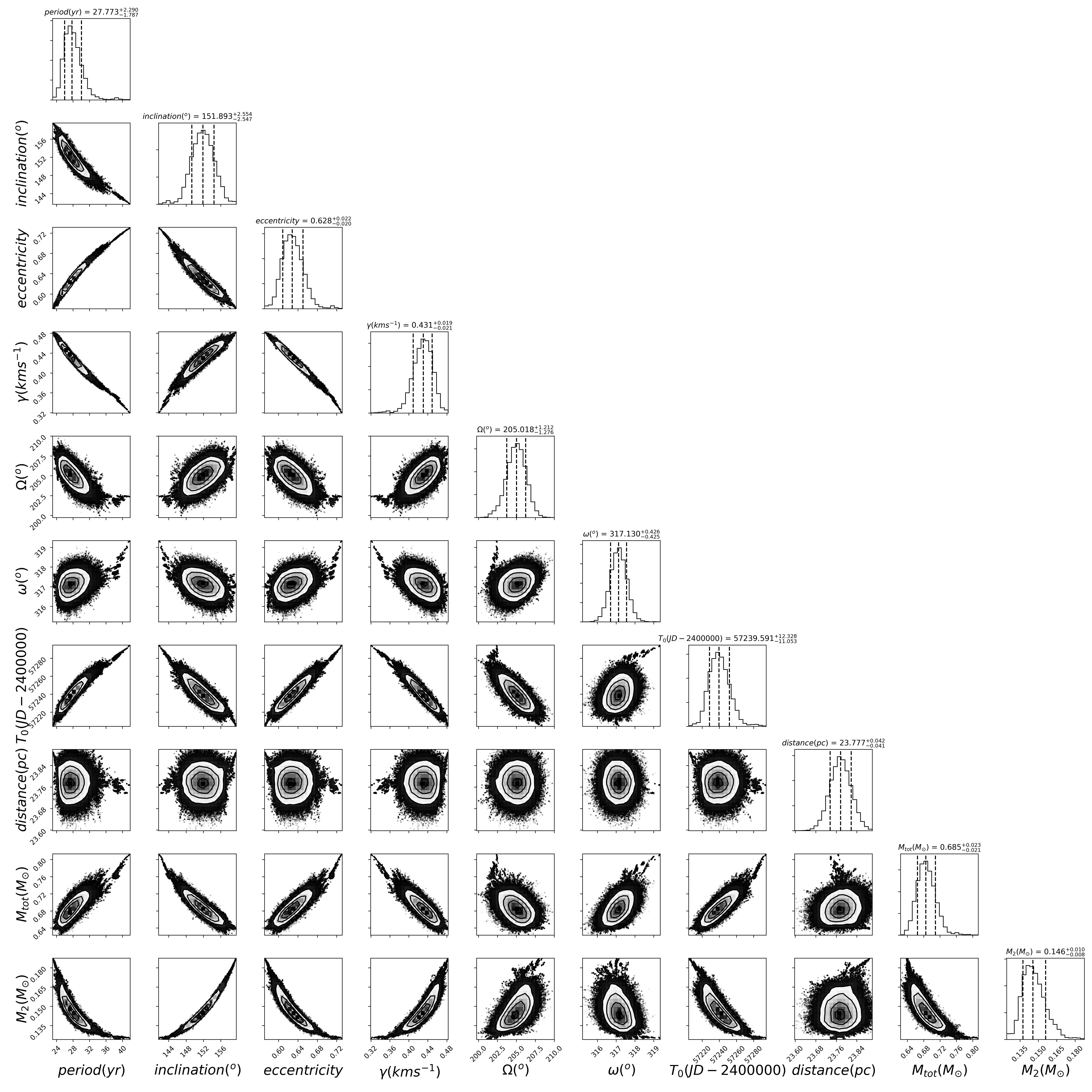}
\caption{The corner plot for PT-MCMC fit to the orbit of HIP 113201AB, incorporating direct imaging data and radial velocity data in the orbital fit.
}
\label{fig:fit_corner_HIP113201_dirvonly}       
\end{figure*}

\begin{figure*}
\includegraphics[width=0.55\textwidth]{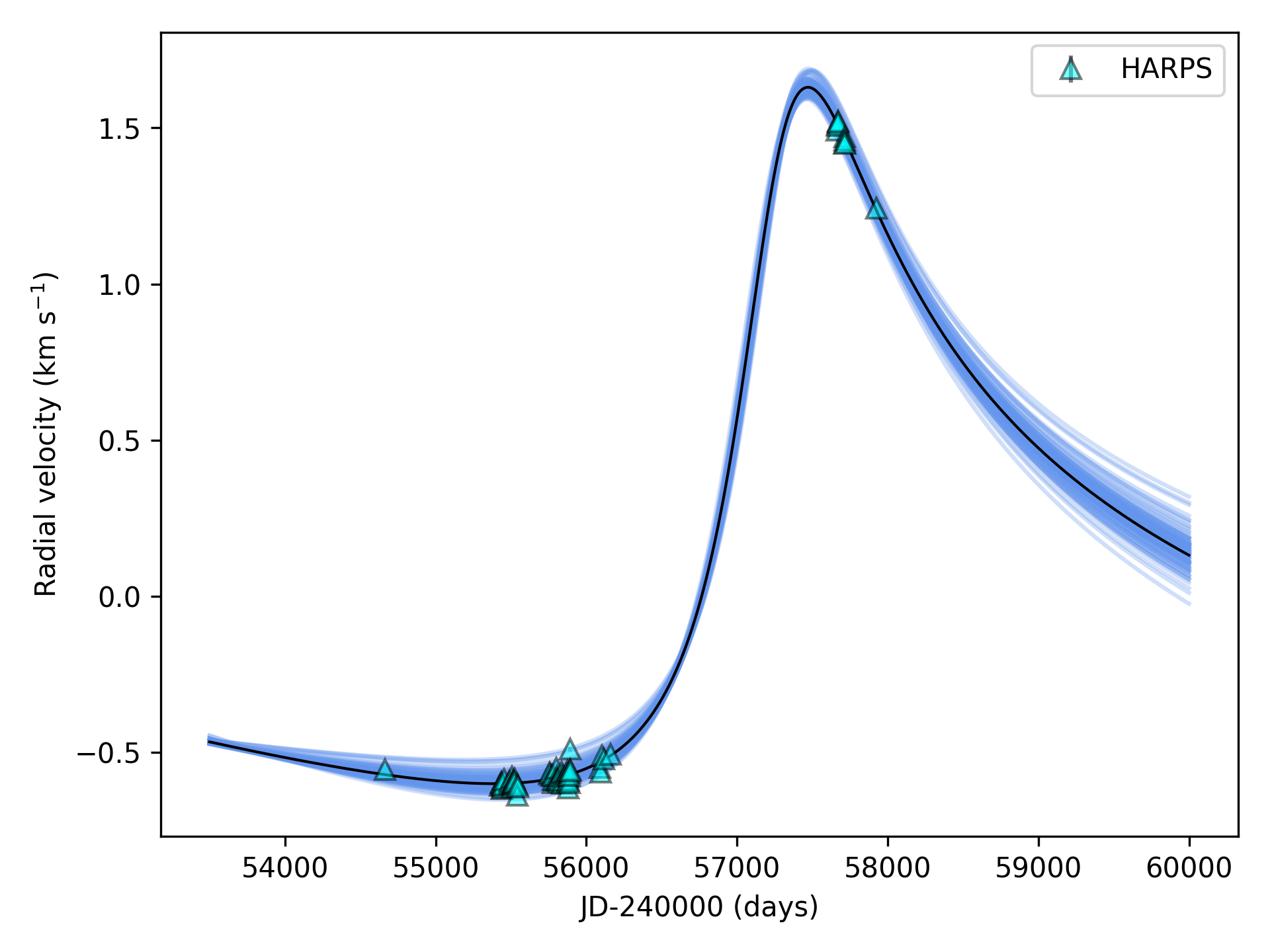}
\includegraphics[width=0.45\textwidth]{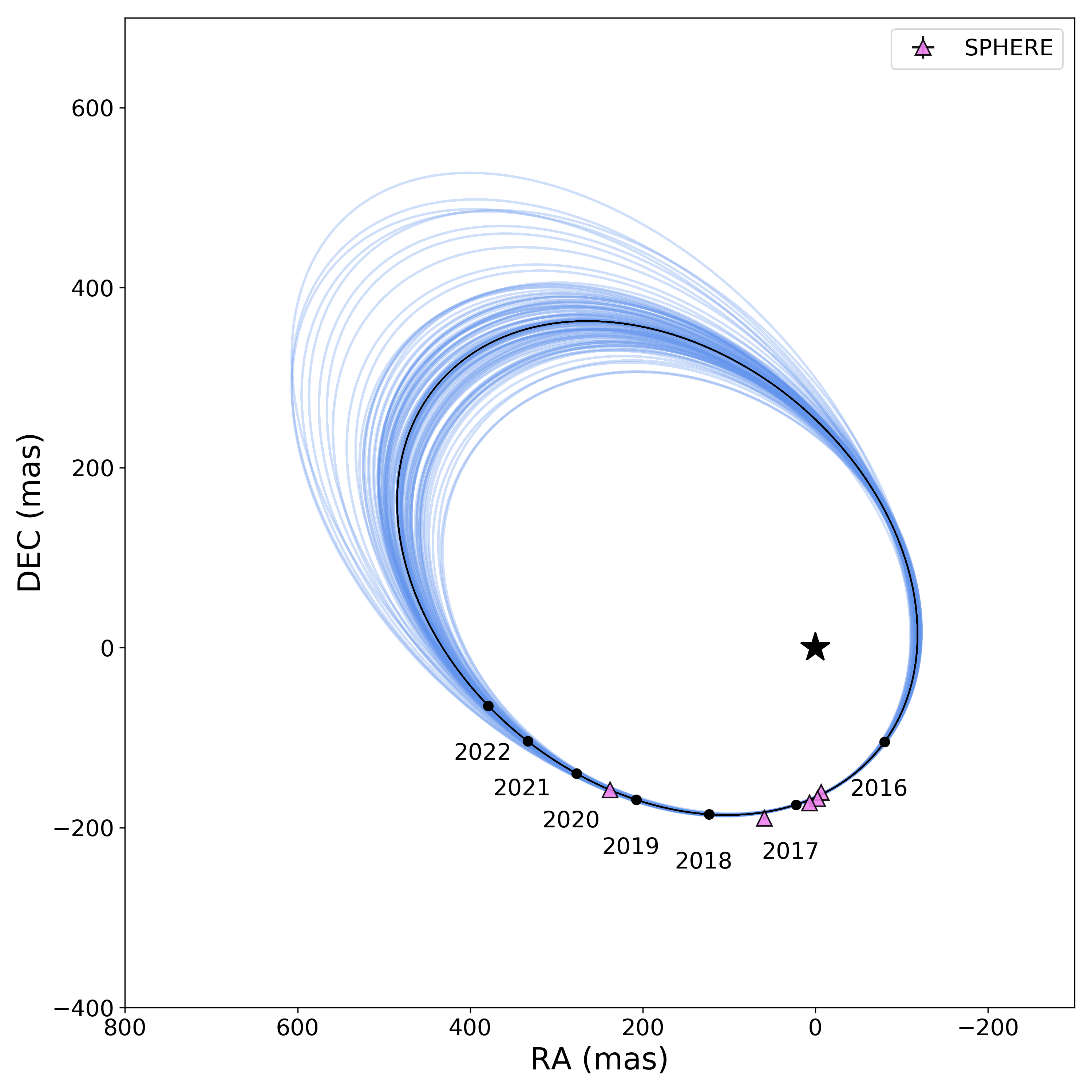}
\caption{{Left:} Radial velocity data vs. model comparison for HIP 113201AB.  HARPS data points are plotted as cyan triangles.  The best fit orbit to the direct imaging data and radial velocity data is plotted as a solid black line; blue lines depict 100 random orbits taken from the final converged PT-MCMC posterior pdf. {Right:} Direct imaging data vs. model comparison for HIP 113201AB.  SPHERE data points are plotted as lavender triangles.  The best fit orbit to the direct imaging data and radial velocity data is plotted as a solid black line; blue lines depict 100 random orbits taken from the final converged PT-MCMC posterior pdf.  The position of the primary is depicted with a black star symbol.
}
\label{fig:RV_HIP113201_dirvonly}       
\end{figure*}

\begin{figure*}
\includegraphics[scale=0.5]{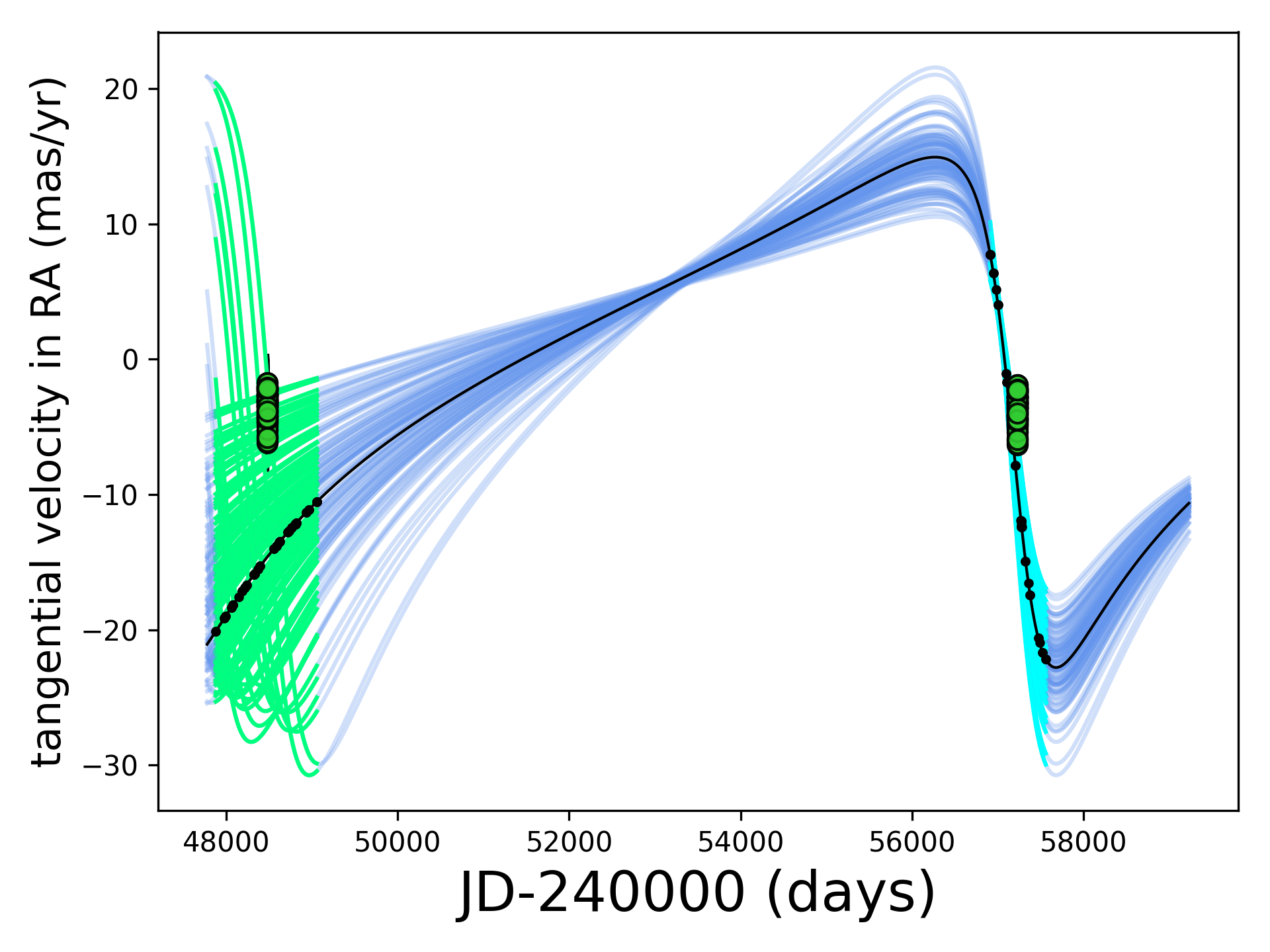}
\includegraphics[scale=0.5]{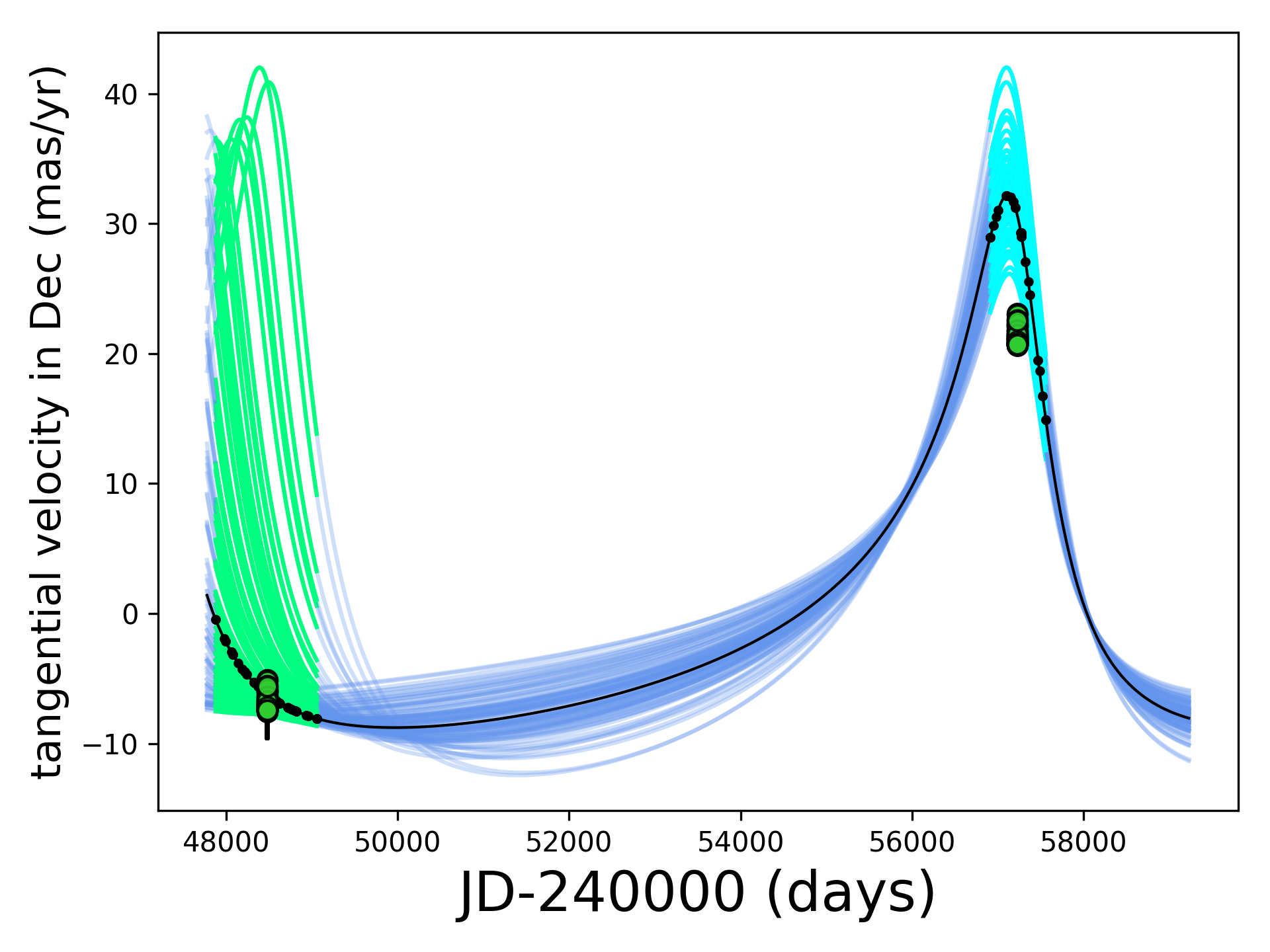}
\caption{Model tangential velocity compared to Hipparcos and Gaia barycenter-corrected proper motion anomalies for HIP 113201AB. The tangential velocity in RA is plotted in the left panel and the tangential velocity in Declination is plotted in the right panel. The best fit orbit to the direct imaging data and radial velocity data is plotted as a solid black line; blue lines depict 100 random orbits taken from the final converged PT-MCMC posterior pdf.  Hipparcos and Gaia barycenter-corrected proper motion anomalies are plotted as green points for the same 100 random orbits; because the barycenter correction depends on primary and secondary mass, these vary slightly depending on the orbit selected.  The Hipparcos mission lifetime for the 100 random orbits is highlighted in green; the Gaia DR2 observation period is highlighted in cyan.  The small circle points depict the dates at which Hipparcos and Gaia measurements were acquired.  
}
\label{fig:fit_tangential_HIP113201_dirvonly}       
\end{figure*}

Corner plots for all parameters of this fit are presented in Fig.~\ref{fig:fit_corner_HIP113201_dirvonly}.  The best fit orbit, as well as 100 orbits randomly selected from the posterior probability distribution function are plotted alongside the RV and direct imaging data in Fig.~\ref{fig:RV_HIP113201_dirvonly}.  The best fit orbit and 100 randomly selected orbits from the posterior of the model tangential motion on the sky compared to the barycenter-corrected Hipparcos and DR2 proper motion anomalies are presented in Fig.~\ref{fig:fit_tangential_HIP113201_dirvonly}.

\FloatBarrier

\section{HIP 113201 orbital fits -- direct imaging, RV, and Hipparcos/Gaia proper motion anomalies \label{app:HIP113201_dirvhipgaia}}

\begin{figure*}
\includegraphics[width=\textwidth]{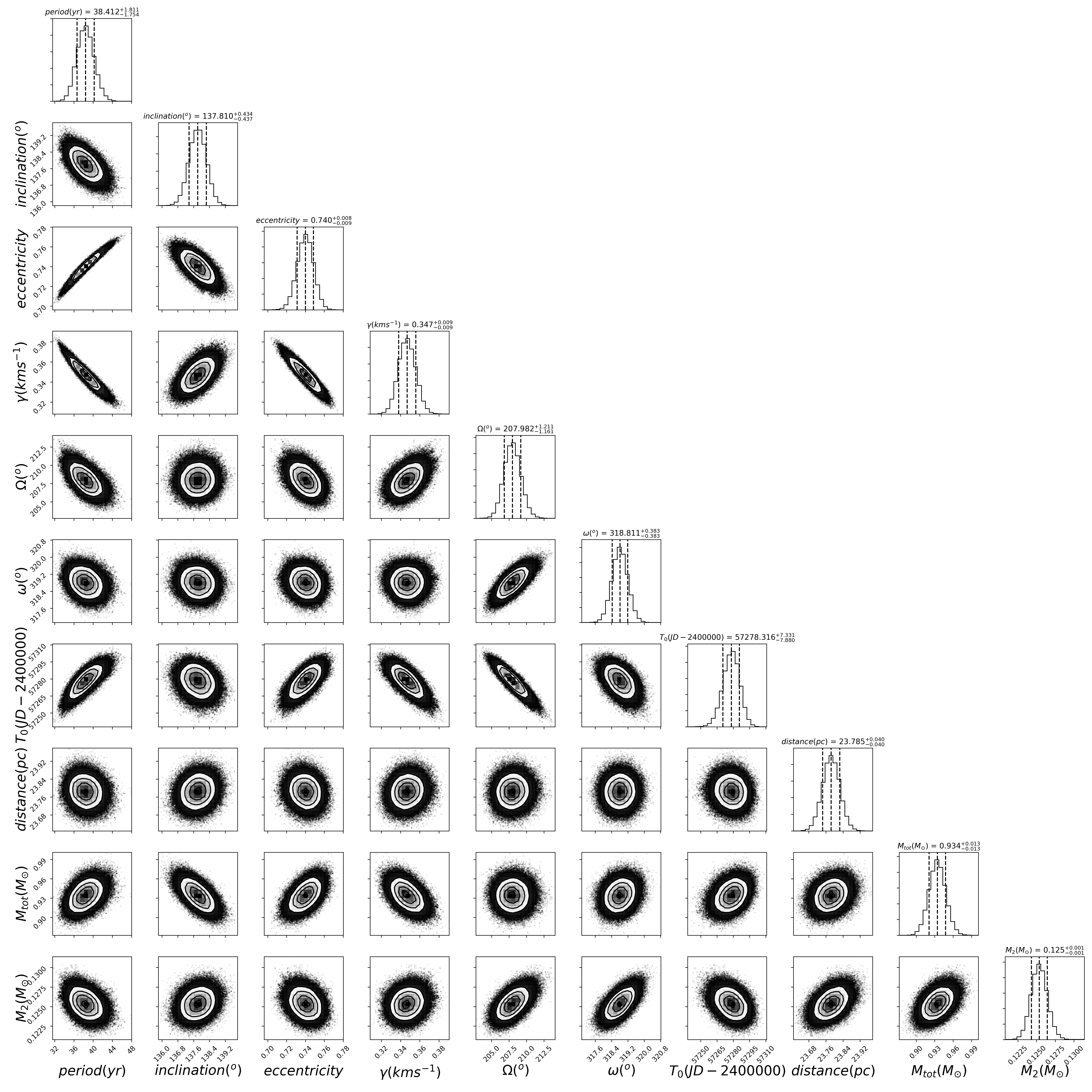}
\caption{The corner plot for PT-MCMC fit to the orbit of HIP 113201AB, incorporating direct imaging, radial velocity and both Hipparcos and Gaia proper motion anomalies in the orbital fit.}
\label{fig:fit_corner_HIP113201_hg}       
\end{figure*}

\begin{figure*}
\includegraphics[width=0.55\textwidth]{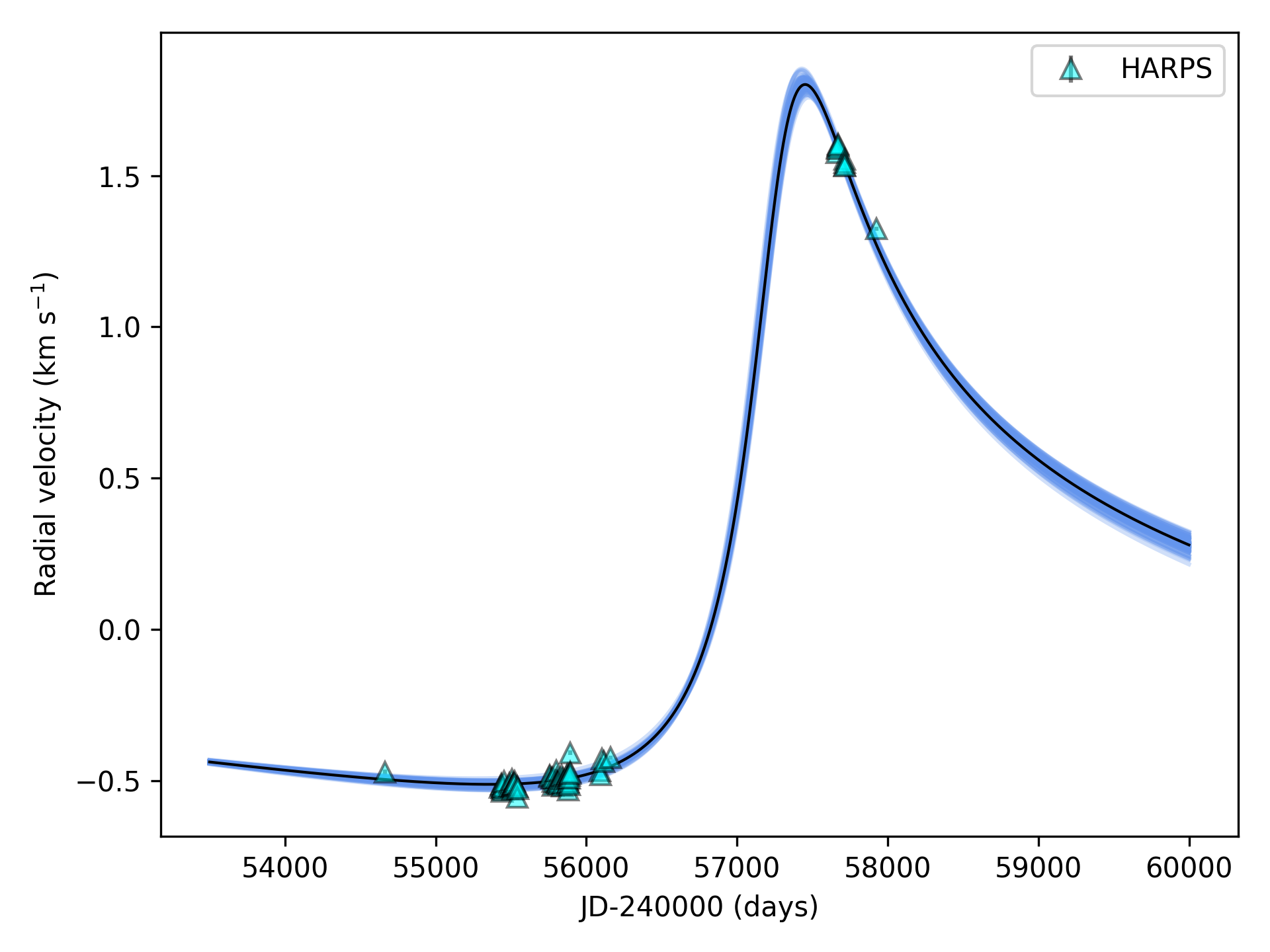}
\includegraphics[width=0.45\textwidth]{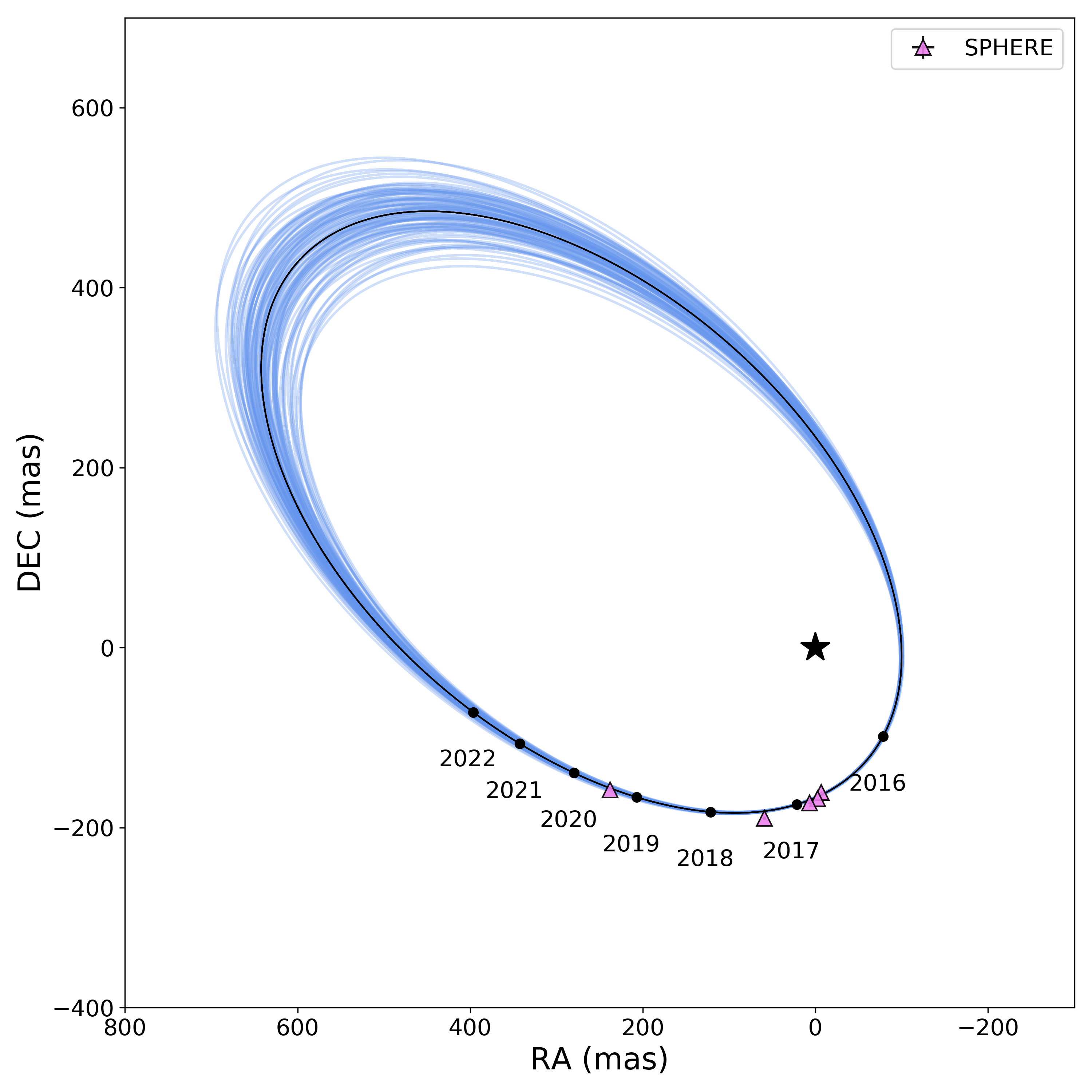}
\caption{\textit{Left:} Radial velocity data vs. model comparison for HIP 113201AB.  HARPS data points are plotted as cyan triangles.  The best fit orbit to the direct imaging data, radial velocity data and both Hipparcos and Gaia proper motion anomalies is plotted as a solid black line; blue lines depict 100 random orbits taken from the final converged PT-MCMC posterior pdf. \textit{Right}: Direct imaging data vs. model comparison for HIP 113201AB.  SPHERE data points are plotted as lavender triangles.  The best fit orbit to the direct imaging data, radial velocity data and both Hipparcos and Gaia proper motion anomalies is plotted as a solid black line; blue lines depict 100 random orbits taken from the final converged PT-MCMC posterior pdf.  The position of the primary is depicted with a black star symbol. 
}
\label{fig:RV_HIP113201_hg}       
\end{figure*}

\begin{figure*}
\includegraphics[scale=0.5]{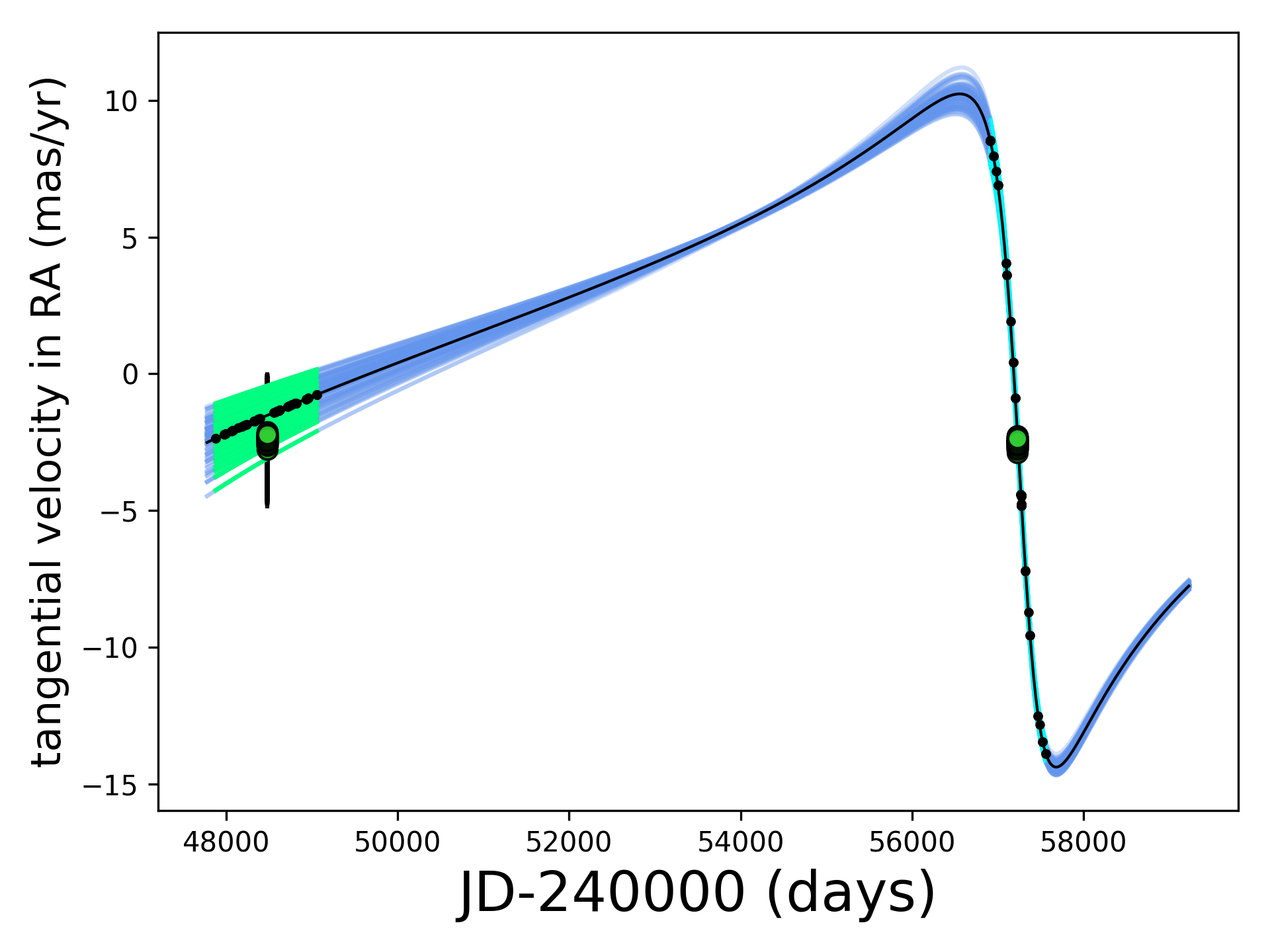}
\includegraphics[scale=0.5]{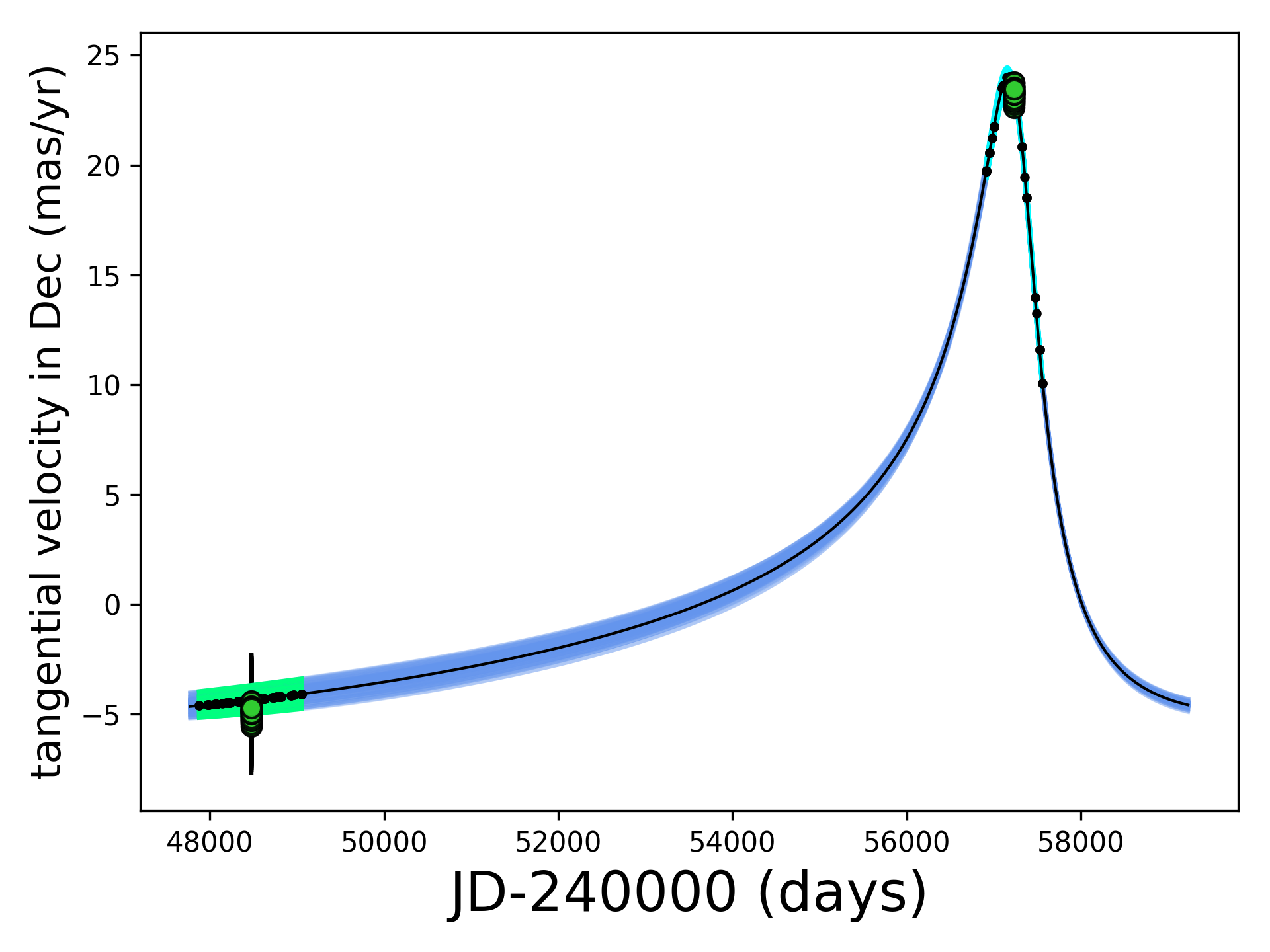}
\caption{Model tangential velocity compared to Hipparcos and Gaia barycenter-corrected proper motion anomalies for HIP 113201AB. The tangential velocity in RA is plotted in the left panel and the tangential velocity in Declination is plotted in the right panel.  The best fit orbit to the direct imaging data, radial velocity data and both Hipparcos and Gaia proper motion anomalies is plotted as a solid black line; blue lines depict 100 random orbits taken from the final converged PT-MCMC posterior pdf.  Hipparcos and Gaia barycenter-corrected proper motion anomalies are plotted as green points for the same 100 random orbits; because the barycenter correction depends on primary and secondary mass, these vary slightly depending on the orbit selected.  The Hipparcos mission lifetime for the 100 random orbits is highlighted in green; the Gaia DR2 observation period is highlighted in cyan.  The small circle points depict the dates at which Hipparcos and Gaia measurements were acquired.  
}
\label{fig:fit_tangential_HIP113201_hg}       
\end{figure*}

Corner plots for all parameters of this fit are presented in Fig.~\ref{fig:fit_corner_HIP113201_hg}.  The best fit orbit, as well as 100 orbits randomly selected from the posterior probability distribution function are plotted alongside the RV and direct imaging data in Fig.~\ref{fig:RV_HIP113201_hg}.  The best fit orbit and 100 randomly selected orbits from the posterior of the model tangential motion on the sky compared to the barycenter-corrected Hipparcos and DR2 proper motion anomalies are presented in Fig.~\ref{fig:fit_tangential_HIP113201_hg}.

\FloatBarrier

\FloatBarrier

\section{HIP 113201 orbital fits -- direct imaging, RV, and Gaia proper motion anomaly \label{app:HIP113201_dirvgaia}}

\begin{figure*}
\includegraphics[width=\textwidth]{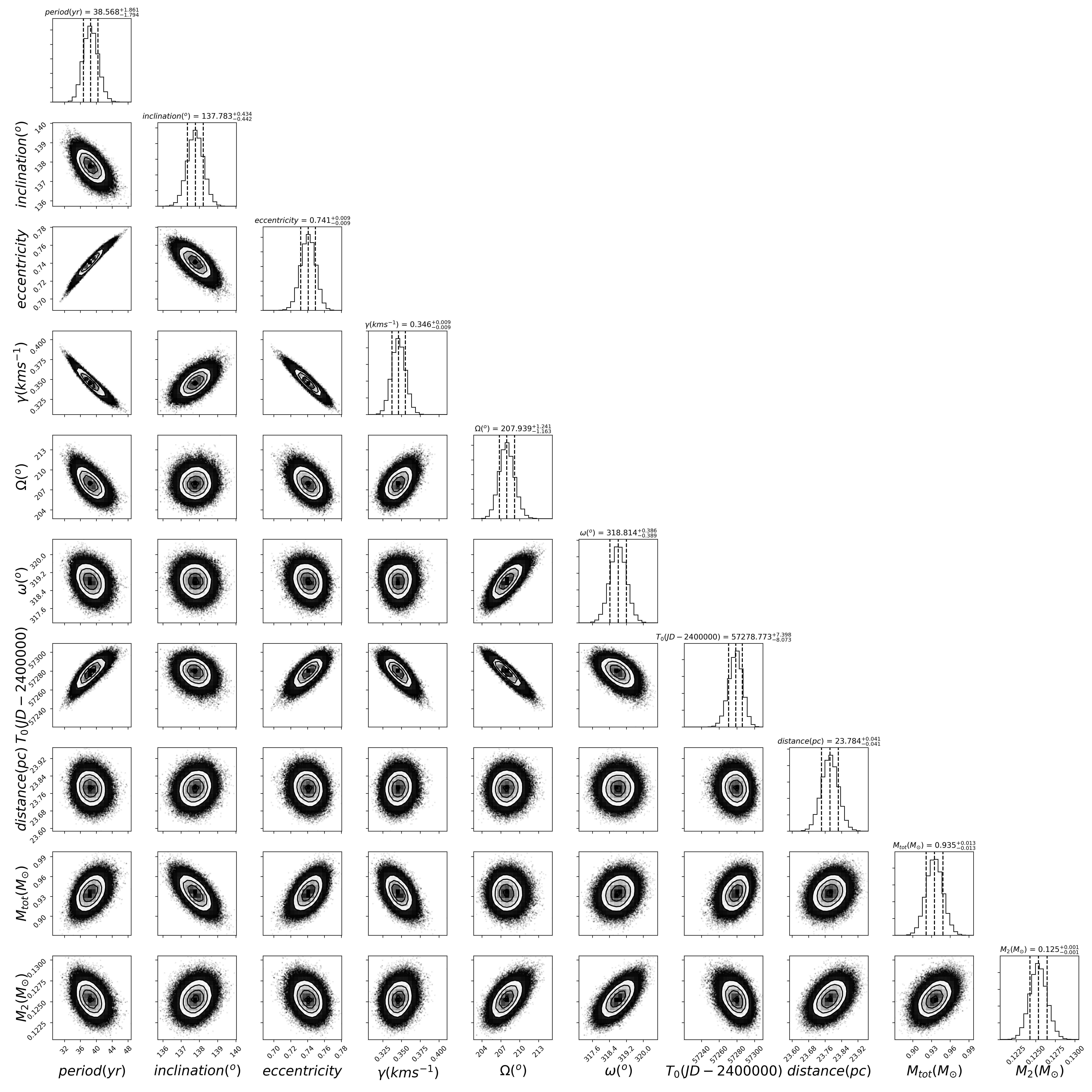}
\caption{The corner plot for PT-MCMC fit to the orbit of HIP 113201AB, incorporating direct imaging data, radial velocity data and the Gaia
proper motion anomaly in the orbital fit.}
\label{fig:fit_corner_HIP113201_g}       
\end{figure*}

\begin{figure*}
\includegraphics[width=0.55\textwidth]{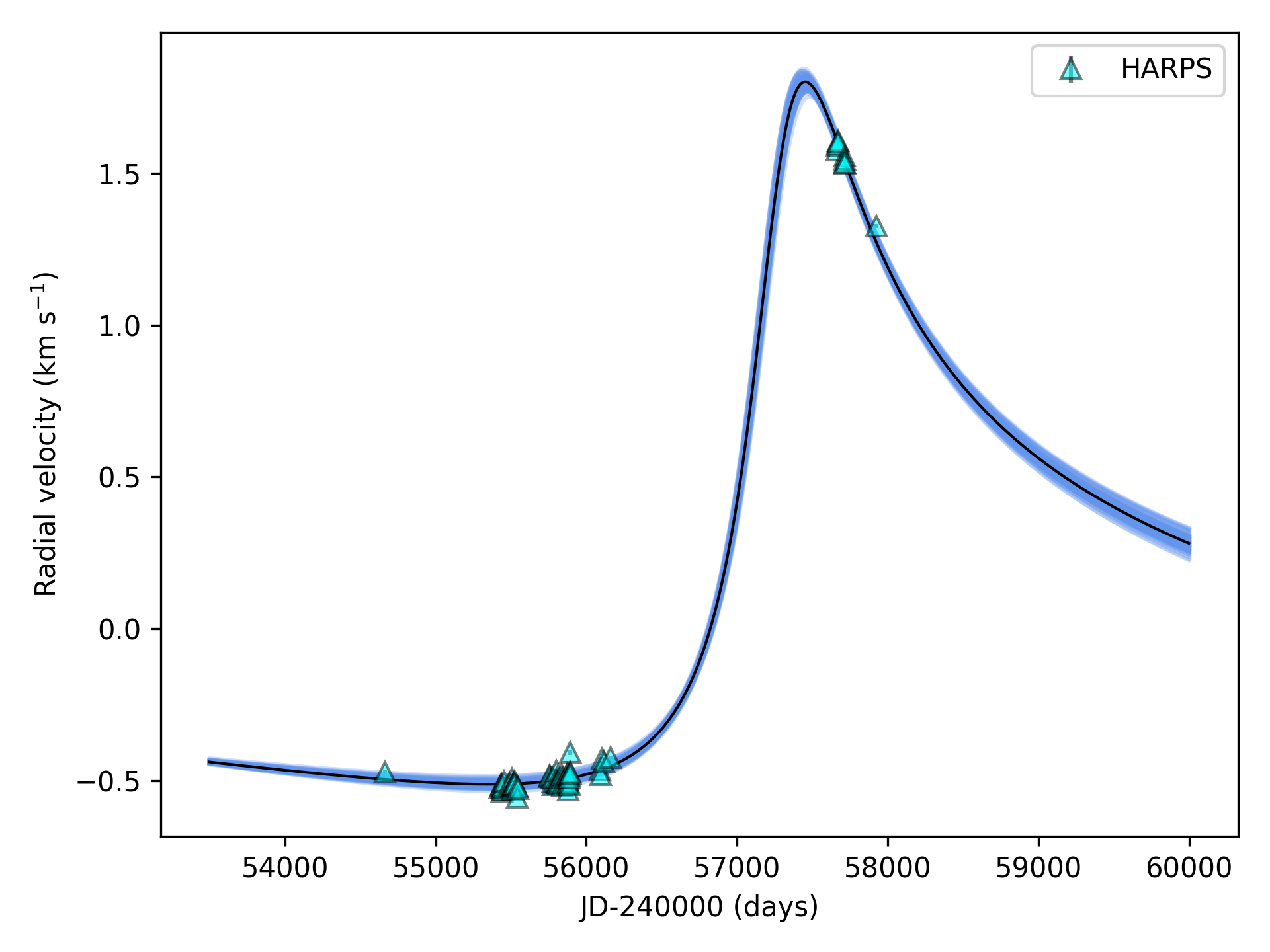}
\includegraphics[width=0.45\textwidth]{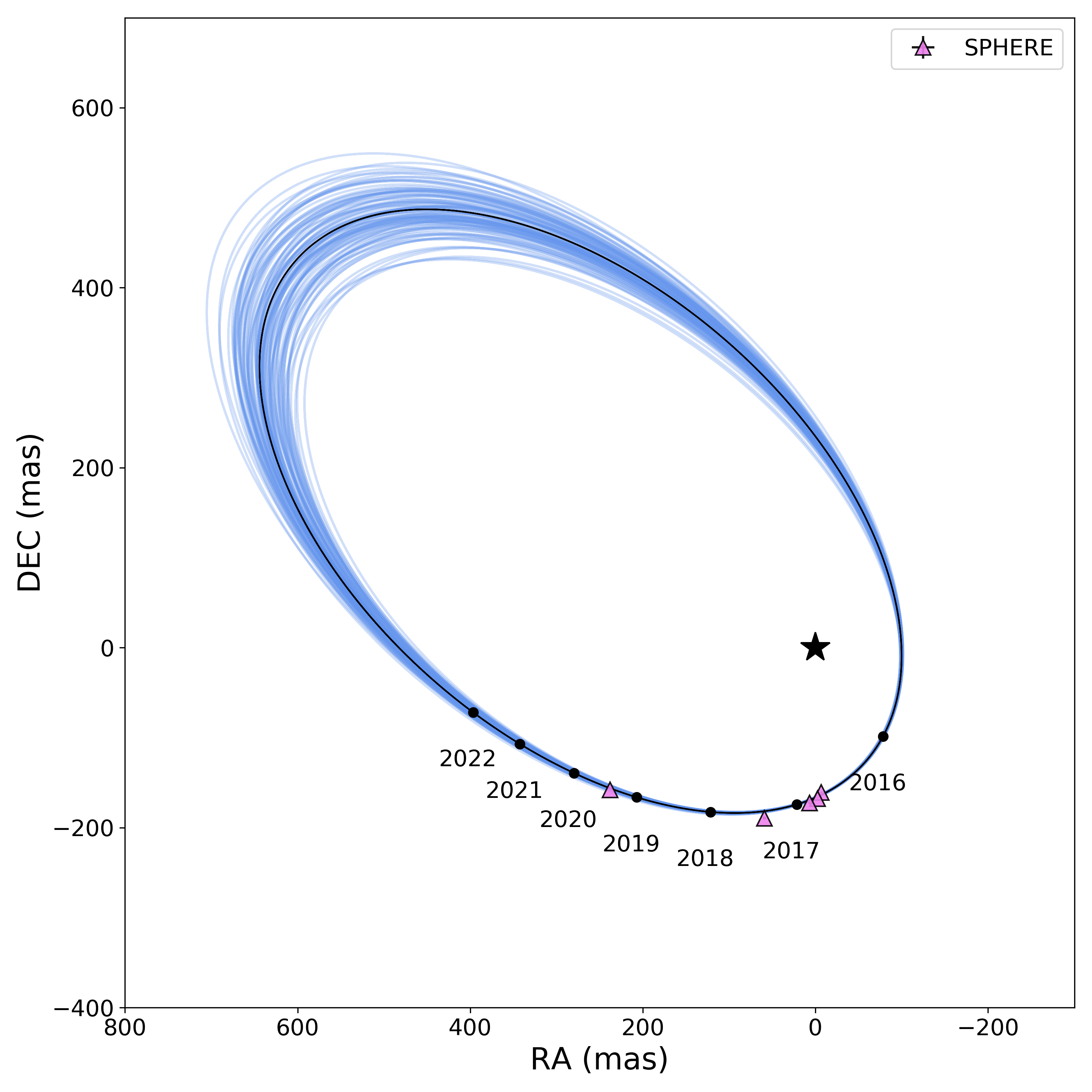}
\caption{{Left:} Radial velocity data vs. model comparison for HIP 113201AB.  HARPS data points are plotted as cyan triangles.  The best fit orbit to the direct imaging data, radial velocity data and Gaia proper motion anomaly is plotted as a solid black line; blue lines depict 100 random orbits taken from the final converged PT-MCMC posterior pdf. {Right:} Direct imaging data vs. model comparison for HIP 113201AB.  SPHERE data points are plotted as lavender triangles.  The best fit orbit to the direct imaging data, radial velocity data and Gaia proper motion anomaly is plotted as a solid black line; blue lines depict 100 random orbits taken from the final converged PT-MCMC posterior pdf.  The position of the primary is depicted with a black star symbol.
}
\label{fig:RV_HIP113201_g}       
\end{figure*}

\begin{figure*}
\includegraphics[scale=0.5]{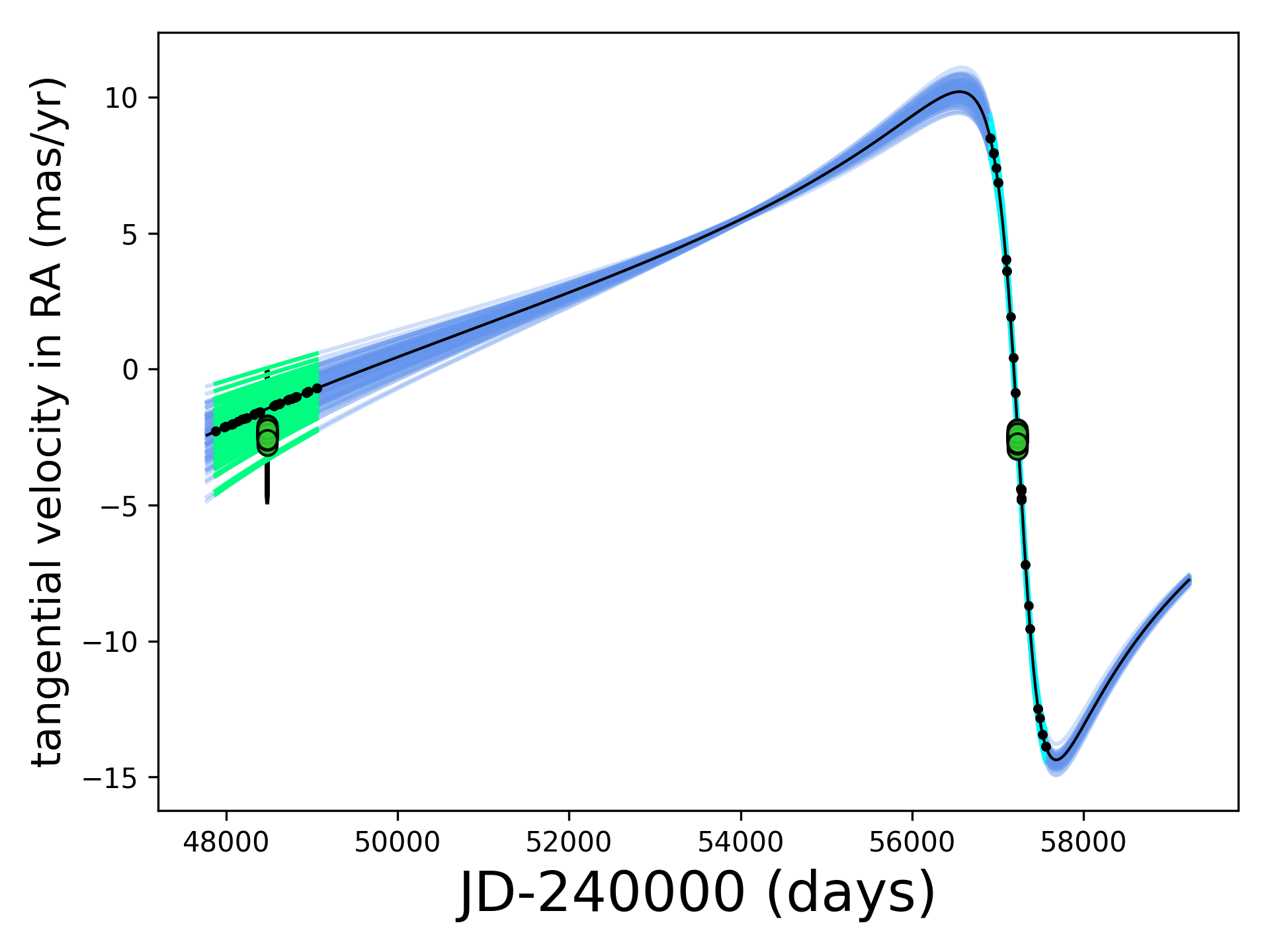}
\includegraphics[scale=0.5]{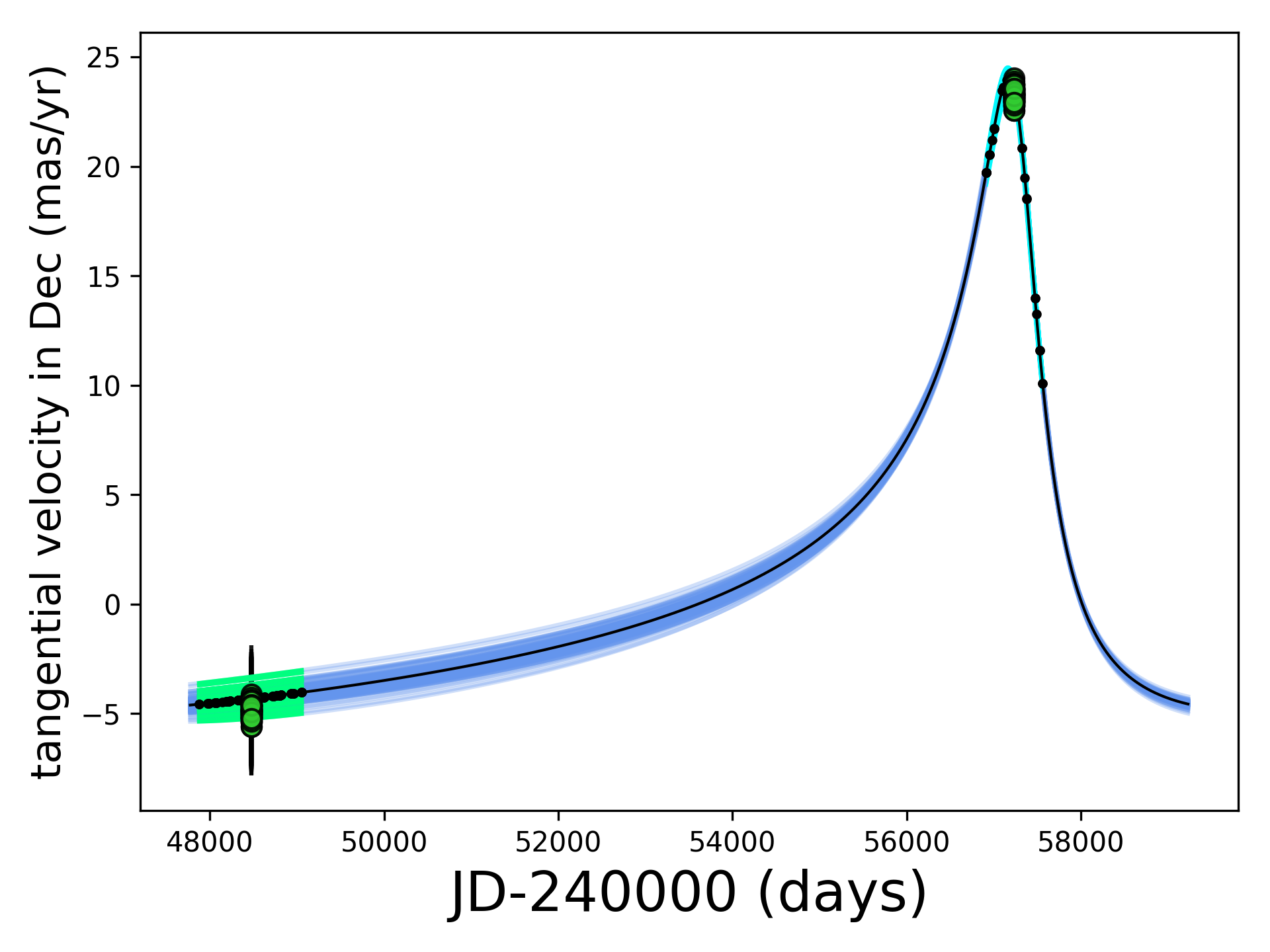}
\caption{Model tangential velocity compared to Hipparcos and Gaia barycenter-corrected proper motion anomalies for HIP 113201AB.  The tangential velocity in RA is plotted in the left panel and the tangential velocity in Declination is plotted in the right panel.  The best fit orbit to the direct imaging data, radial velocity data and Gaia proper motion anomaly is plotted as a solid black line; blue lines depict 100 random orbits taken from the final converged PT-MCMC posterior pdf.  Hipparcos and Gaia barycenter-corrected proper motion anomalies are plotted as green points for the same 100 random orbits; because the barycenter correction depends on primary and secondary mass, these vary slightly depending on the orbit selected.  The Hipparcos mission lifetime for the 100 random orbits is highlighted in green; the Gaia DR2 observation period is highlighted in cyan.  The small circle points depict the dates at which Hipparcos and Gaia measurements were acquired.  
}
\label{fig:fit_tangential_HIP113201_g}       
\end{figure*}

Corner plots for all parameters of this fit are presented in Fig.~\ref{fig:fit_corner_HIP113201_g}. 
The best fit orbit, as well as 100 orbits randomly selected from the posterior probability distribution function are plotted alongside the RV and direct imaging data in Fig.~\ref{fig:RV_HIP113201_g}.  The best fit orbit and 100 randomly selected orbits from the posterior of the model tangential motion on the sky compared to the barycenter-corrected Hipparcos and DR2 proper motion anomalies are presented in Fig.~\ref{fig:fit_tangential_HIP113201_g}.

\end{appendix}

\end{document}